\documentclass[journal]{IEEEtran}
\ifCLASSINFOpdf
  \usepackage[pdftex]{graphicx}
\else
\fi

\usepackage{amsmath}
\usepackage{cases}
\usepackage{multicol}
\usepackage{multirow}
\usepackage{subcaption}
\usepackage[dvipsnames]{xcolor}
\usepackage[normalem]{ulem}
\usepackage{listings}
\usepackage{textcomp}
\usepackage[ruled, vlined]{algorithm2e}
\usepackage{url}
\usepackage{flushend}

\ifCLASSOPTIONcompsoc
  \usepackage[nocompress]{cite}
\else
  \usepackage{cite}
\fi

\begin{document}

\title{Freezer: A Specialized NVM Backup Controller for Intermittently-Powered Systems}

\author{Davide Pala,
        Ivan Miro-Panades,~\IEEEmembership{Member,~IEEE},
        and~Olivier Sentieys,~\IEEEmembership{Member,~IEEE}
\IEEEcompsocitemizethanks{\IEEEcompsocthanksitem D. Pala and O. Sentieys are with Univ. Rennes, Inria, Rennes,
France.\protect
\IEEEcompsocthanksitem I. Miro-Panades is with Univ. Grenoble Alpes, CEA List, Grenoble, France.}
\thanks{Manuscript received April 10, 2020; revised xxx.}}



\maketitle

\begin{abstract}
The explosion of IoT and wearable devices determined a rising attention towards energy harvesting as source for powering these systems. 
In this context, many applications cannot afford the presence of a battery because of size, weight and cost issues. 
Therefore, due to the intermittent nature of ambient energy sources, these systems must be able to save and restore their state, in order to guarantee progress across power interruptions.
In this work, we propose a specialized backup/restore controller that dynamically tracks the memory accesses during the execution of the program.
The controller then commits the changes to a snapshot in a Non-Volatile Memory (NVM) when a power failure is detected. 
Our approach does not require complex hybrid memories and can be implemented with standard components. 
Results on a set of benchmarks show an average $8\times$ reduction in backup size. Thanks to our dedicated controller, the backup time is further reduced by more than $100\times$, with an area and power overhead of only 0.4\% and 0.8\%, respectively, w.r.t. a low-end IoT node.
\end{abstract}


\IEEEpeerreviewmaketitle


\section{Introduction}\label{sec:introduction}
In the context of IoT, many applications cannot afford the presence of a battery because of size, weight and cost issues.
The recent advancement in the Non-Volatile Memory (NVM) technologies is paving the way for Non-Volatile Computing Systems.
These systems are able to sustain computations under unstable power, by quickly saving the state of the full system in a non-volatile fashion.
Thus, Non-Volatile Processors (NVPs) may allow battery-less designs without suffering from frequent power losses inherent in energy harvesting scenarios.

In related work, both software- and hardware-level solutions were proposed to cope with the backup and restore problem.
Software-based approaches are implemented on platforms that include both some SRAM and an addressable NVM used to store the backup, as the one presented in \cite{zwerg_82_2011}.
Checkpoints are placed at compile time~\cite{ransford_mementos:_2011}. 
Then, at run-time the supply voltage is checked and, if an imminent power failure is identified ($V_{dd} < V_{th}$), a backup of the stack and the registers is executed.
In some works, backups are only executed when a power failure interrupt is triggered and the full volatile state (SRAM and registers) is copied to the NVM~\cite{balsamo_hibernus:_2015, balsamo_hibernus_2016}.
Other approaches do not take advantage of the volatile SRAM and exploit the NVM as the only system memory, backing-up only the registers in the event of a power outage~\cite{jayakumar_quickrecall:_2015, choi_achieving_2019}.
Software-level solutions can be implemented on available hardware, but they normally come with a big overhead in terms of both backup time and energy. 

Hardware solutions on the other hand usually implement fully Non-Volatile Processors (NVP).
NVPs mostly make use of emerging NVM technologies to implement complex hybrid memory elements (nvFF and nvSRAM, non-volatile registers, and SRAM memory, respectively) that allow for very fast parallel backup and restore operations \cite{yu_non-volatile_2011, wang_3us_2012, liu_4.7_2016, wang_a_130nm_feram_2017, sakimura_10.5_2014, senni_non-volatile_2016}. 
However, introducing these hybrid memory elements is intrusive. Moreover, it usually comes with a significant area overhead and often results in increased delay and active power.
Additional limitations on the amount of data that can be saved and restored in parallel is imposed by the peak current consumption required to drive all the NVM bit cells at the same time.
To mitigate these problems, distributed small non-volatile arrays, where groups of flip-flops are backed-up in sequence, are proposed in \cite{bartling_8mhz_2013}.
An adaptive restore controller for configuring the parallelism of the nvSRAM restore operation, trading off peak current with restore speed is instead presented in~\cite{liu_4.7_2016}. 

The use of NVM enables persistence across power failures but it also introduces the problem of consistency for the data stored in the NVM \cite{ransford_nonvolatile_2014}.
To address the consistency issue and improve reliability of the system, a software framework that performs a copy-on-write of modified pages of the NVM in a shadow memory area is developed in~\cite{choi_achieving_2019}.
The consistency problem can be also addressed via static analysis or with hardware techniques \cite{liu_lightweight_2016}.
In particular, hybrid nvFFs can be used in a hardware scheme where an enhanced store buffer is used to treat the execution of stores to the NVM as speculative, until a checkpoint is reached~\cite{liu_lightweight_2016}.
Two counters are also used to periodically trigger checkpoints based on the number of executed stores or on the number of executed instructions.
Previous work has also focused the attention to the problem of optimal checkpoint placement, as in \cite{ghodsi_optimal_2017} where online decisions on checkpoints are taken based on a table filled offline using Q-learning.

In this paper, we propose Freezer, a hardware backup and restore controller that is able to reduce the amount of data that needs to be backed-up.
Our approach avoids the high cost of hardware fully NVP architectures since it can be implemented with plain CMOS technology.
Furthermore, contrary to other hardware based approaches such as non-volatile processors \cite{liu_4.7_2016, sakimura_10.5_2014, ma_architecture_2015}, our proposed controller is a component that can be integrated in existing SoCs, without requiring modification of the processor architecture.
Moreover, Freezer achieves better performance than pure software approaches.
Our contributions can be summarized as follows:
\begin{itemize}
    \item We propose an analysis of different backup strategies based on the use of memory access traces. 
    \item We introduce an oracle based backup strategy that provides the optimal lower bound for the backup size.
    \item We present a hardware backup controller, Freezer, that dynamically keeps track of the changes in the program state and commits these changes in the NVM before the power failure. 
    The controller spies the address signal of the SRAM and uses dirty bits to track modified addresses with a block granularity.
    \item We conduct an analysis of the trade-offs and a design space exploration for our proposed strategy. Results on a set of benchmarks show an average $8\times$ reduction in backup size. Thanks to Freezer, the backup time is further reduced by more than $100\times$, with a very low area and power overhead.
    \item We compare the memory access energy of three different system architectures: SRAM+NVM, NVM-only and cache+NVM, showing that NVM-only systems take on average $3.74\times$ to $3.35\times$ more energy than SRAM+NVM with full-memory backup and $6.19\times$ to $4.22\times$ more when compared to Freezer. Our strategy shows a clear advantage also when compared to cache+NVM architecture, requiring in average $7.8\times$ and $5.9\times$ less energy, with respectively RRAM and STTRAM as main memory.
\end{itemize}

The rest of the paper is organized as follows. 
In Section \ref{sec:context}, we present some background information and related works. 
In Section \ref{sec:sys-modeling} we describe the main system models and architectures for a transiently powered device, and we present a model for evaluating the memory access energy of different system architectures.
In Section \ref{sec:backup-model}, we introduce and discuss the model for the backup strategies.
Section \ref{sec:traces} explains how the memory access traces of the benchmarks are processed and analysed.
Section \ref{sec:freezer} presents Freezer backup controller, its algorithm, and some area and power synthesis results.
We report several comparison results of our study in Section \ref{sec:results}.
Finally, we briefly discuss our approach and draw the conclusions in Sections \ref{sec:discussion} and \ref{sec:conclusion}.

\section{Background and Related Work}\label{sec:context}
In this section, we briefly present the context around non-volatile processors and the problem of state retention in energy harvesting applications.
We then present the motivation from which this paper is derived.

In related work, both software- and hardware-level solutions were proposed to guarantee forward progress across unpredictable power failures.
There are two main approaches to cope with the backup and restore problem:
periodic check-pointing \cite{ransford_mementos:_2011, choi_achieving_2019},  and
on-demand backup \cite{balsamo_hibernus:_2015, balsamo_hibernus_2016, jayakumar_quickrecall:_2015}.
Periodic check-pointing systems try to guarantee forward progress by repeatedly executing some check-pointing tasks, interleaved with the computation. 
These check-points are usually placed by the compiler, according to some heuristic.
At run-time, when a check-point is reached, the system decides if a backup should be executed.
In \cite{ransford_mementos:_2011}, for example, the supply voltage level is checked to determine whether there is enough energy or if a snapshot should be taken. 
After a power outage, the state will be rolled back to the last saved state and the execution will resume from the last check-point that was reached. 
This approach has the advantage that backup size can be optimised, as the location of each check-point is known in advance.
In \cite{choi_achieving_2019}, checkpoints are instead taken based on the expiration of a timer, but only the registers are saved as the system uses only NVM as its main memory. To avoid consistency issues with NVM updates happening between a checkpoint and a power failure, the modified NVM pages are saved with a copy-on-write mechanism on a shadow memory area.
These periodic check-pointing techniques also introduce overhead due to the execution of unnecessary checkpoints and backups, moreover they may lead to the re-execution of part of the code after the rollback.

\begin{figure*}[htbp]
    \centering
    \includegraphics[width=.85\linewidth]{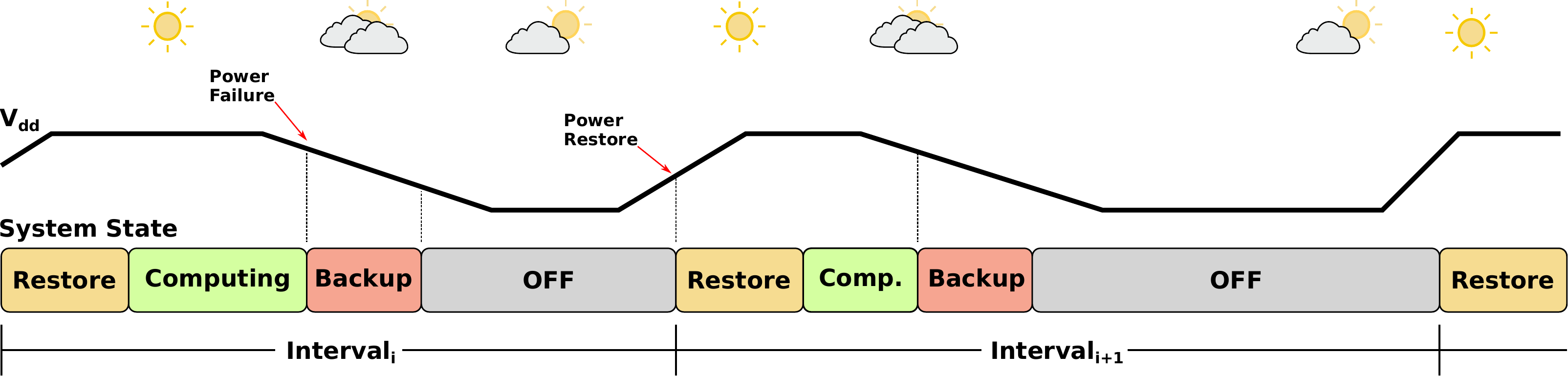}
    \caption{Division of execution time in intervals and system state during an interval.}
    \label{fig:interval}
\end{figure*}

On-demand backup tries to avoid the run-time overheads introduced with periodic check-pointing by waiting until a power failure is detected before executing the backup. 
The typical behavior of an on-demand backup system is depicted in Fig. \ref{fig:interval},
which shows how the system responds to a power failure, signaled by a decrease in the supply voltage (Vdd), by interrupting the computation and by entering in the \textit{Backup} phase.
When the backup is completed, the system goes in the \textit{OFF} state, where it will wait until the power resumes.
When the power is newly available, the platform can leave the \textit{OFF} state and start the recovery.
The new interval begins when the system enters the \textit{Restore} phase, to recover the state saved in the previous backup.
When the restore is completed the system can resume the computation.

Some hardware-based solutions can also be considered as implementation of on-demand backups.
As an example, in \cite{su_a_ferroelectric_2017}, the non-volatile processor is paired with a dedicated voltage detector used to trigger the backup mechanism.
The main disadvantage with these techniques is that they often require a full backup of the system memory, as it is difficult to know in advance when a power failure will happen and thus saving only the required memory is complicated. 

To mitigate this problem some offline static analysis technique have been proposed \cite{zhao_software_2015, zhao_stack-size_2017}. 
In particular, in \cite{zhao_stack-size_2017}, an offline analysis of the code is used to find the backup positions that reduce the stack size.
These positions are marked in the code with the insertion of special label instructions. 
At run-time a dedicated hardware module will wait for the power failure signal. After this signal, the execution continues until the
program reaches the label instruction. Then, this dedicated hardware module executes the backup.
These techniques require a compile-time analysis, with a detailed energy model of the platform.
Moreover they tend to introduce overhead as they need to modify the program code \cite{zhao_software_2015} and the internal architecture of the processor \cite{zhao_stack-size_2017}.

Non-volatile processors can also be considered implementation of on-demand backup, as they focus on having very fast backup (and restore) in response to power failure.
In \cite{ma_architecture_2015}, architectures and techniques for implementing non-pipelined, pipelined, and out-of-order (OoO) non-volatile processors are proposed.
The proposed techniques try to optimise the backup size of the internal state of the processor, using techniques such as dirty bits for a selective backup of the register file.
Contrary to our approach these architectures rely on NVM or hybrid memories for the persistence of the main memory.
Moreover these techniques are in general very intrusive, as they require an in depth modification of the internal architecture of the processor. 

To address the problem of full memory backup in an on-demand scheme, we propose a hardware backup controller, Freezer, that is able to optimise the size of the backup based on the information collected at run-time.
Our proposed controller is an independent component that can be integrated in existing SoCs, without requiring changes to the internal architecture of the processor core.

In this work, we focus on how to optimise the backup of the main memory and we do not consider the problem of saving the internal state of the processor.
However the state of the CPU could be managed via software by the processor by copying its internal register into the main memory before starting the back up.
Other techniques are proposed in the literature to save the internal registers.
Common hardware-based solutions use nvFFs based on different technologies, such as STTRAM~\cite{sakimura_10.5_2014}, MRAM-based nvFFs~\cite{senni_non-volatile_2016}, FeRAM~\cite{su_a_ferroelectric_2017}, ReRAM~\cite{liu_4.7_2016}, and the use of FeRAM distributed mini arrays~\cite{bartling_8mhz_2013} or the use of nvFFs and NVM blocks for the backup of internal registers~\cite{ma_architecture_2015}.

\section{System Modelling}
\label{sec:sys-modeling}
\subsection{Considered System Model}\label{sec:sys-model}

\begin{figure}[ht]
    \centering
    \includegraphics[width=.9\linewidth]{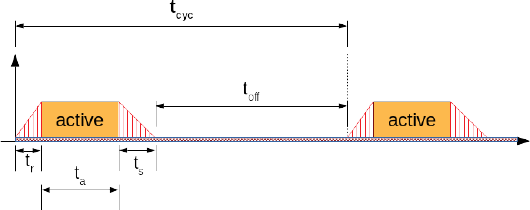}
    \caption{Detail of an execution cycle between two consecutive power outages.}
    \label{fig:energy-cycle} 
\end{figure}

Energy harvesting is seen as a promising source to power future battery-less IoT systems.
However, due to the unpredictable nature of the energy source, these systems will be subject to sudden power outages.
This could cause the execution of program to be unexpectedly interrupted.
Thus, in these intermittently (or transiently) powered systems, the execution is divided in multiple power cycles, i.e., intervals, as shown in Fig. \ref{fig:interval}.
The timing break-down of one of these intervals is depicted in Fig. \ref{fig:energy-cycle}.
$t_{cyc}$ is the duration of this on-off cycle and is defined as $t_{cyc}=t_r+t_a+t_s+t_{off}$, where $t_a$ is the time in active state where the system is executing some software tasks, $t_{off}$ the time in the power-off state, and $t_r$, $t_s$ the time to restore, save (backup) the data from, to the NVM, respectively.
The energy consumed by the system during $t_{cyc}$ can be modelled as (adapted from \cite{hager_a_scan-chain:2017})
\begin{equation}
    E_{c} = E_{s} N_{s} + E_{r} N_{r} + P_{on} t_{a} + P_{off} t_{off},
    \label{eq:energy}
\end{equation}
where $E_{s}$ and $E_{r}$ are the energy required respectively for saving and restoring one word, $N_s$ and $N_r$ the total number of words to save and restore.
$P_{on}$ and $P_{off}$ are the power consumed during the active state and off state, respectively.
In this type of intermittently-powered systems, usually $P_{off}$ is zero as the state is retained in a non-volatile manner, thus the all system including the processor core can be fully shut-down.
Moreover, $N_s$ and $N_r$ are usually equal and often coincide with the full size of the volatile system state~\cite{balsamo_hibernus:_2015}.

Considering an \textit{on-demand backup} system that only performs a backup before a power failure, the total execution time $t_{exec}$ of a program can be modeled as (adapted from \cite{balsamo_hibernus:_2015})
\begin{equation}
    t_{exec} = t_{prog} + n_i \times (t_s + t_r + \overline{t_{off}}),
    \label{eq:time}
\end{equation}
where $t_{prog}$ is the time needed for running the whole program without interruptions, $n_i$ the number of interruptions, $t_s$ and $t_r$ the save and restore time, respectively, and $\overline{t_{off}}$ the average off time. 
 
Our approach, Freezer, aims at reducing the size of the backup ($N_{s}$), thus also reducing $t_s$ and the total execution time and backup energy. 
Moreover, the hardware implementation of our approach guarantees an additional decrease to the backup and restore time and energy, by eliminating the overhead due to software operations.

In this paper, we assume that the system has a reliable way to detect a power failure and we also assume that the system has enough power to complete the backup.
Therefore we do not investigate the problem of how to deal with incomplete backup.
To have a stronger guarantee on the consistency of the system state after recovery a double buffering scheme can be applied, such that a new backup does not overwrite the previous one on the NVM.
Moreover, we do not deal with the issue of how to detect a power failure.
For this problem there are also solutions proposed in the literature, such as dedicated voltage detector \cite{su_a_ferroelectric_2017}.

\subsection{System Architecture} \label{sec:sys-arch}
In the field of non-volatile processors for energy harvesting applications, there are several possible architectural choices for achieving state retention. 
The most common approaches are the following:
\begin{itemize}
    \item A CPU with an SRAM and an addressable NVM.
        The NVM might serve as a backup of the full memory space of the SRAM but might also be addressable by the processor. 
    \item A CPU with an SRAM and a backup-only NVM or a CPU with a hybrid nvSRAM as in \cite{liu_4.7_2016}. 
    \item A CPU with an NVM as main system memory as in \cite{sakimura_10.5_2014, wang_a_130nm_feram_2017, jayakumar_quickrecall:_2015, senni_non-volatile_2016, choi_achieving_2019}. 
    \item A CPU with an SRAM-based cache and an NVM as the only system memory \cite{ghodsi_optimal_2017, ma_architecture_2015}. 
\end{itemize}
\begin{figure}
    \centering
    \includegraphics[width=.85\linewidth]{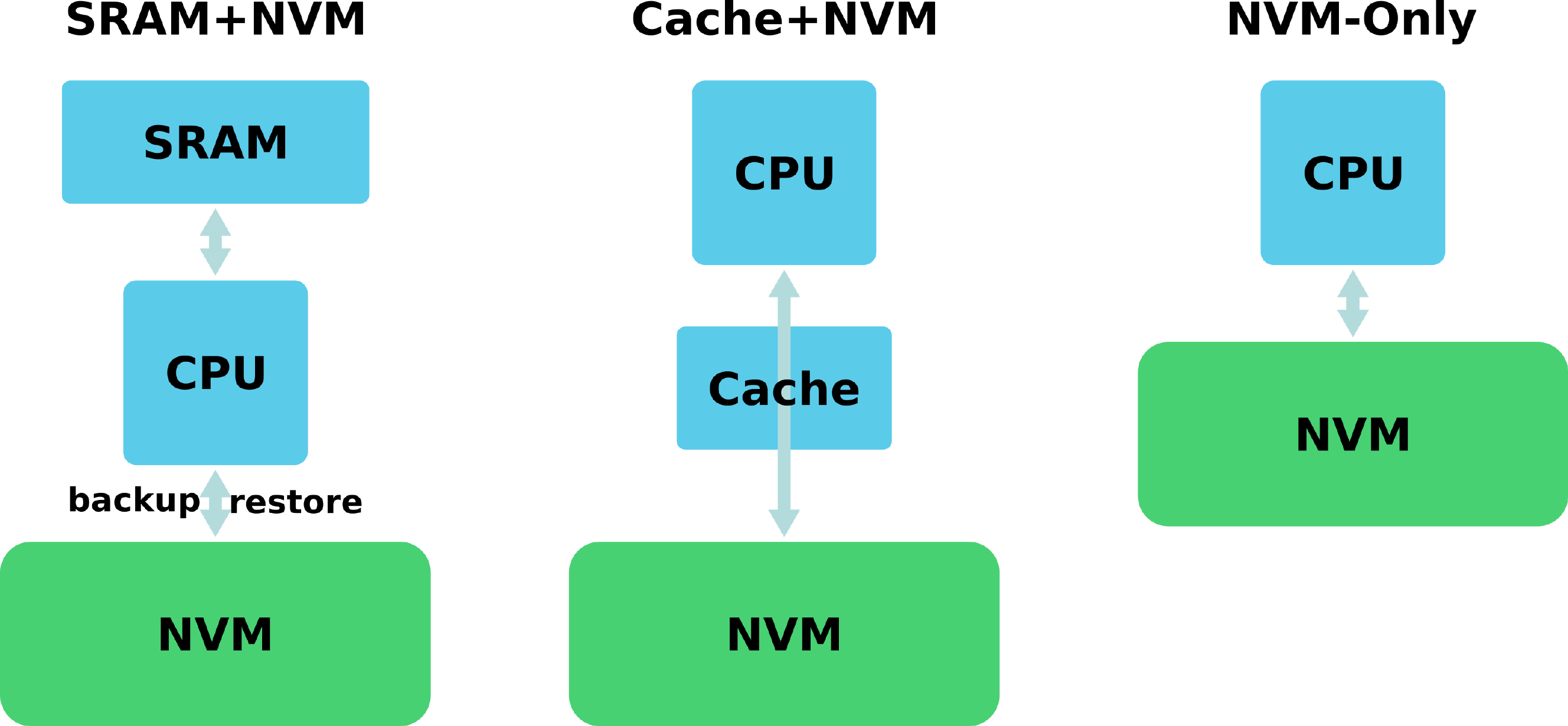}
    \caption{Architectural models for non-volatile state retention.}
    \label{fig:arch_models}
\end{figure}
These approaches can be grouped into the three basic architectures depicted in Fig. \ref{fig:arch_models}. 
The first two approaches have in common the SRAM+NVM architecture, which, as shown in Fig. \ref{fig:arch_models}, exploits SRAM for execution and NVM for enabling backup and restore operations.
The NVM-only approach relies solely on NVM as its main memory.
Cache+NVM uses NVM as the main memory with the addition of a volatile cache.

A common choice for implementing intermittently-powered systems is to use commercially available SoCs with an embedded addressable NVM.
As this NVM is addressable, this type of systems is the common choice for implementing software-based retention schemes~\cite{balsamo_hibernus:_2015, ransford_mementos:_2011}.
Another option explored in related work is that of using hybrid nvSRAM \cite{yu_non-volatile_2011, liu_4.7_2016}.
This choice allows to exploit the main advantages of SRAM (fast read/write and low access power), while also obtaining fast parallel backup through the paired non-volatile memory elements.
This means that the non-volatile elements are not directly accessible by the programmer, instead the non-volatility is made transparent by the hardware.
A conceptually simple solution to guarantee state retention is to exploit only an NVM as the main memory.
This solution is proposed in \cite{sakimura_10.5_2014}, where the system is fully based on STTRAM.
Another example is given by the software approach of QuickRecall \cite{jayakumar_quickrecall:_2015}, where the available SRAM is not used and the system runs only on the FeRAM.
As with hybrid nvSRAM, the non-volatility is transparent to the programmer.
Also, in this case, there is no need to copy the data in the event of a power failure.
In \cite{ma_architecture_2015} methods for the backup and recovery of the internal state are proposed and compared considering non pipelined, pipelined, and out of order (OoO) processor architectures. 
These solutions can also be considered NVM-only type of systems, as they use NVM as their main memory, with the addition of hybrid or NVM caches in the case of the OoO processor.

Unfortunately, some of these new NVM technologies are still immature and often they do not provide the same level of performance in terms of access time and access energy as the SRAM \cite{yu_emerging:2016}.
Moreover, NVM-only designs must also face the issue of wear and the reduced endurance that characterises many of the emerging NVM technologies.
To mitigate this problem, a possible solution could be to use register-based or SRAM-based store buffers. 
As an example, enhanced store buffers are proposed in \cite{liu_lightweight_2016} to postpone the execution of NVM writes, treating store operations as speculative.
Though the limited size of the store buffer still results in very frequent checkpoints and a large number of NVM writes. 
Another possible answer to mitigate NVM writes speed and endurance problem could be to use an SRAM-based cache to buffer the accesses to the main NVM.
Although this type of architecture could be of some interest for higher performance systems, it is not very common in small IoT edge nodes. 
This is because adding a cache would significantly increase both the dynamic and static power consumption during the active period.

\begin{figure}[ht]
\centering
\includegraphics[width=.6\linewidth]{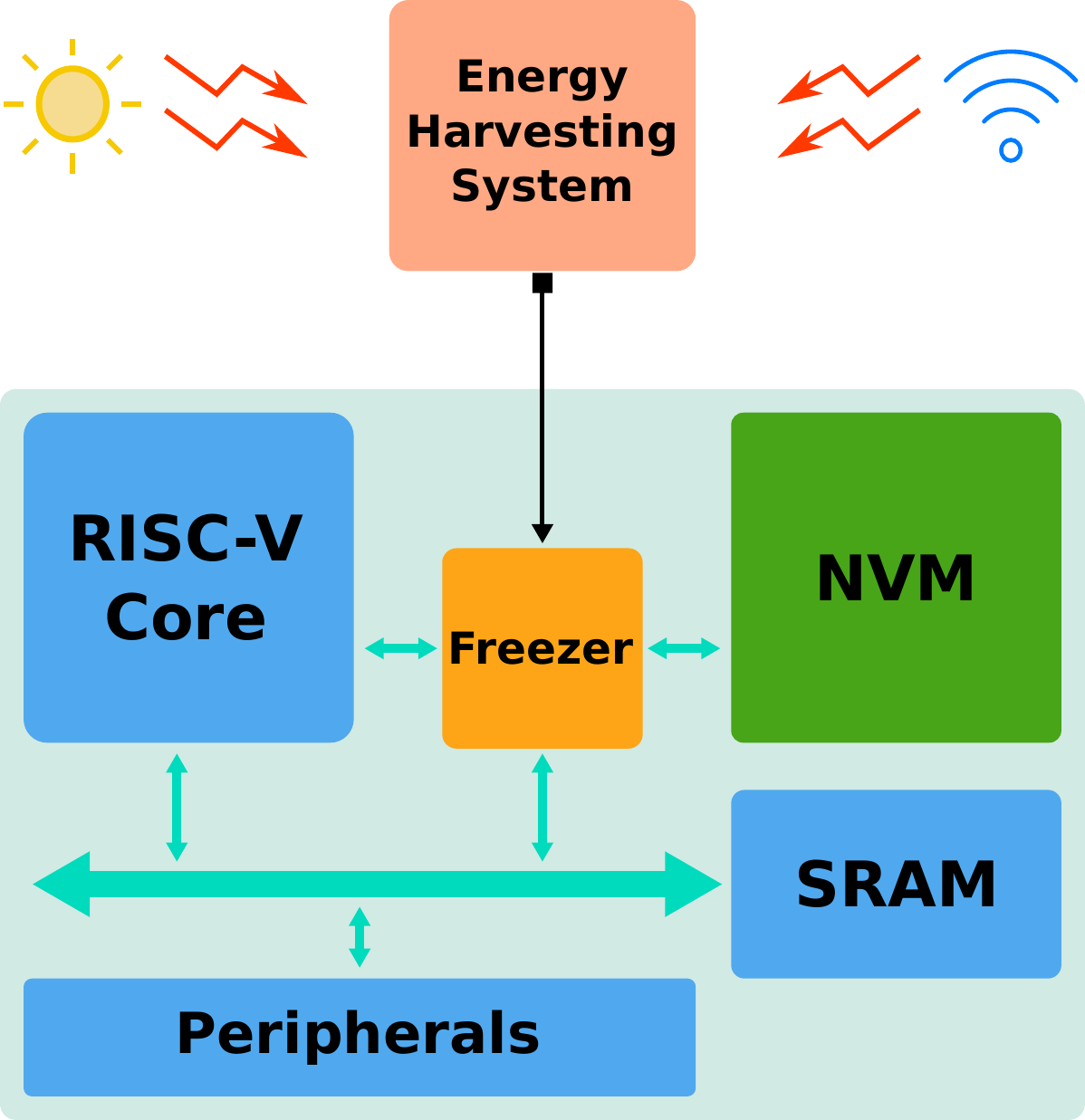}
\caption{General overview of a system implementing Freezer.}
\label{fig:system-arch} 
\end{figure}

In this work, we consider an architecture that comprises a micro-controller with an SRAM as main memory and an NVM that is used by our proposed backup controller, Freezer, to save (and restore) the state of the system before (and after) a power failure. 
The general overview of such architecture is depicted in Fig. \ref{fig:system-arch}.
The micro-controller we consider implements the RISC-V Instruction Set Architecture (ISA).

\subsection{Modelling Memory Access Energy}
\label{sec:res:energy}
The energy required for a backup operation is dominated by the data transfers between the SRAM and the NVM, and will be proportional to the backup size.
This energy will mostly be determined by the write energy of the NVM, that can be even $100\times$ that of the SRAM \cite{yu_emerging:2016}.
Our approach provides a reduction of the backup energy by decreasing the number of data transfers and by improving the speed of the process compared with a software based backup strategy. 

In this section, we provide a simplified model to evaluate and compare the energy cost of some of the different system architectures introduced in Section \ref{sec:sys-arch}. Results provided in Section \ref{sec:results} are based on this model.
In particular, we derive the energy cost in terms of memory accesses for the following types of memory models:
\begin{itemize}
    \item SRAM + NVM for backup,
    \item NVM only,
    \item cache + NVM as main memory.
\end{itemize}
For the \textit{SRAM+NVM} architecture, we consider both a system which performs a full memory backup and a system with Freezer. 
For this system, the energy cost associated with memory accesses can be expressed as
\begin{equation}
    E_{SRAM+NVM} = E_{prog} + E_{backup} + E_{restore}
    \label{eq:energy-sram+NVM} 
\end{equation}
where $E_{prog}$ is the energy of the memory accesses needed for running the program. 
\begin{equation}
    E_{prog} = E_{sram/r}  N_{load} + E_{sram/w}  N_{store}
    \label{eq:energy-prog}
\end{equation}
where $E_{sram/r}$ and $E_{sram/w}$ are the read and write energy of the SRAM, and $N_{load}$ and $N_{store}$ are the total number of load and store operations, respectively.
The additional cost required by a platform with both SRAM and NVM are expressed in Eq. \ref{eq:energy-sram+NVM} by the energy for the backup $E_{backup}$ and by the energy for the restore $E_{restore}$, defined respectively as
\begin{equation}
    E_{backup} = N_{s} (E_{sram/r} + E_{nvm/w}),
    \label{eq:energy-backup}
\end{equation}
\begin{equation}
    E_{restore} = N_r (E_{nvm/r} + E_{sram/w}).
    \label{eq:energy-restore}
\end{equation}
The energy for the backup depends on the total size of the backup $N_s$ and on the energy required for reading from SRAM $E_{sram/r}$ and writing to NVM $E_{nvm/w}$.
$N_s$ is the total number of saved words throughout the full execution.
Similarly $E_{restore}$ can be expressed as the energy for a single transfer (read from NVM and write to SRAM) multiplied by the total number of restored words $N_r$.

For the \textit{NVM-only} architecture there is no need to preform backup and restore operations, as everything is already saved in the NVM.
In this case, the memory access energy is given only by the load and store operations performed for running the program.
The energy cost for a purely non-volatile system that uses a NVM as its main memory is estimated by
\begin{equation}
    E_{NVM} = E_{prog NVM} = E_{nvm/r} N_{load} + E_{nvm/w} N_{store}.
    \label{eq:energy-NVM} 
\end{equation}

The \textit{cache+NVM} architecture comprises both an NVM as its main memory, and an SRAM-based cache to reduce the number of accesses to the NVM.
This system uses a write-back cache controller that performs a flush of the dirty lines on NVM in case of a power failure.
On a cache system, for every operation, the TAG memory is first read to verify if the required address is on the cache or not, then in case of a miss a read from NVM is executed.
Moreover, simultaneous TAG and DATA memory reads are performed inside the cache to sustain high throughput. 
Finally, multiple data words may be accessed in parallel on N-way set-associative cache where only one word is useful. 
Therefore, the energy per read/write operation of this system is much higher that the one with tightly coupled memory (SRAM+NVM).
The energy cost for a cache+NVM system is therefore estimated by
\begin{equation}
    E_{cache} = E_{hits} + E_{misses} + E_{flushes}
    \label{eq:energy-cache}
\end{equation}
where $E_{hits}$ is the energy due to cache hits, $E_{misses}$ the energy penalty due to misses, and $E_{flushes}$ the energy consumed with flushes.
The first part of the energy cost $E_{hits}$ is 
\begin{equation}
    E_{hits} = N_{hit/r}E_{hit} + N_{hit/w}(E_{hit} + E_{cache/w})
    \label{eq:energy-hits}
\end{equation}
where $N_{hit/r}$ and $N_{hit/w}$ are respectively the number of read and write hits, $E_{hit}$ the energy for a single cache access and $E_{cache/w}$ the energy for a write operation inside the cache.
$E_{hits}$ therefore includes the energy due to read hits $N_{hit/r}E_{hit}$ and the energy due to write hits $N_{hit/w}(E_{hit} + E_{cache/w})$.
$E_{misses}$, the energy due to the misses, is expressed as
\begin{equation}
    \begin{aligned}
        E_{misses} &= N_{miss}(E_{miss} + (E_{nvm/r} + E_{cache/w}) \times 8) \\
                   &+ N_{evict} E_{nvm/w}
    \end{aligned}
    \label{eq:energy-misses}
\end{equation}
where $N_{miss}$ is the total number of misses, $E_{miss}$ the energy for a missing access, $N_{evict}$ the total number of evicted words, $E_{nvm/r}$ and $E_{nvm/w}$ are the energy for reading and writing a word in the NVM.
Eq. \eqref{eq:energy-misses} shows that each miss causes the reading of a full block (8 words in our case) from the NVM.
Moreover a missing access may also cause the eviction of a block from the cache resulting in writes to the NVM.
$E_{flushes}$ is caused by the backup of the dirty lines before a power failure happens. This operation requires to scan all the cache lines and write back the dirty ones and is repeated before every power failure.
\begin{equation}
    E_{flushes} = N_i N_{lines}E_{hit} + N_{flush} E_{nvm/w} \times 8 
    \label{eq:energy-flushes}
\end{equation}
where $N_{lines} E_{hit}$ represents the energy for reading all the blocks of the cache and $N_i$ the number of interruptions.
The energy due to the writes to NVM is expressed by $N_{flush}$, the total number of flushed blocks throughout all power failures, multiplied by the energy for writing $8$ words to NVM.


\section{Modeling of the Backup Strategies}\label{sec:backup-model}
By analyzing the memory access sequences, we can identify different backup strategies. 
The \emph{Full Memory Backup} strategy corresponds to the state of the art. 
In this paper, we propose four backup strategies defined as \emph{Used Address (UA)}, \emph{Modified Address (MA)}, and \emph{Modified Block (MB)}, a block-based evolution of the two previous strategies.
The last strategy presented is an \emph{Oracle} and cannot be implemented in a real system as it requires knowledge of the future. 
This oracle is however very useful for comparison, as it gives the optimal lower bound for the backup size. 
In the rest of the paper, a \emph{word} is defined as a 32-bit data.

\subsection{Full Memory Backup}
The first and simplest solution is to backup the full content of the memory at the end of each interval as it is proposed in~\cite{balsamo_hibernus:_2015}.
For our study and fair comparison, we considered a slightly improved version of this strategy that saves only the data section of the program in pages of $512$ bytes (128 words), thus not saving the full memory every time.
As an example, if a program needs a $1000$-byte data space, $1024$ bytes ($2$ pages) will be saved in the NVM.
With this approach, the backup size is a constant for all the intervals, equivalent to the number of pages to be saved.

\subsection{Used Address Backup}
The first strategy that we propose is the \textit{Used Address (UA)} strategy.
UA consists of keeping track of all the different addresses that are accessed (reading and writing) during an interval.
When a power failure is detected every address that was accessed during that interval is saved in the NVM.
In the UA case, only the memory locations that were used during the interval are going to be backed-up.

\subsection{Modified Address Backup}
If the initial snapshot of the program is stored in the NVM, the UA scheme can be improved by implementing the \textit{Modified Address (MA)} backup strategy.
MA only keeps track of the memory locations that are  modified (written) during a power cycle.
Then, before a power outage, only the words that were modified (write operation) are saved to the NVM.
In practice this means saving only the addresses accessed by a store operation at least once during the interval.
This number of addresses gives the size of the backup at the end of the interval. 
It may happen that the data written during execution do not modify the content of the memory. However, to keep the technique simple, we do not track the content of the memory but only the addresses where a write operation happens.

\subsection{Oracle}
The \textit{Oracle} is defined as the strategy that saves only the words that are \textit{alive}.
An address is considered alive when it is going to be read at least once in any future interval.
In other words, a written data is considered alive if it is read in any future intervals, before being modified by any other write at the same address.
A word that will be overwritten before being read is not considered alive and thus is not backed-up by the oracle. 
\begin{figure}[htb]
    \centering
    \includegraphics[width=.9\linewidth]{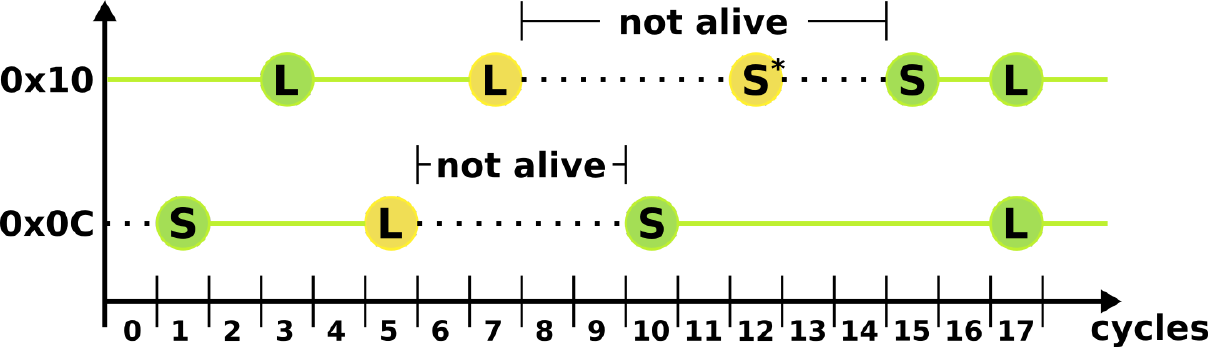}
    \caption{Example of the aliveness of two addresses with Load (L) and Store (S) instructions. The continuous green line indicates that the address is alive. The black dotted line is used when the address is not alive. 
    The store on address \textit{0x10} at cycle 12 (S*) does not make the address alive because it is followed by another store at cycle 15, that overwrites the value written by S*.
    }
    \label{fig:alive-ex}
\end{figure}
Fig. \ref{fig:alive-ex} shows an example of two addresses changing between the \textit{alive} and the \textit{not alive} state as the execution progresses.
In the example, address \textit{0x0C} stops being alive after it is used by the load in cycle 5 and stays \textit{not alive} for the period between the sixth and the ninth clock cycles.
This happens because the Oracle knows that the value will be overwritten by the store executed at clock cycle 10.
Therefore, between clock cycles 6 and 9, it does not consider \textit{0x0C} as an alive address.
For the same reason, address \textit{0x10} stops being alive after the load in cycle 7 and is \textit{not alive} in the time between cycle 8 and cycle 14.
The store operation happening at cycle 12 does not change the state of the address because it is going to be followed by another store instruction that will discard this temporary update.

The \textit{Oracle}, before the power failure, only saves the words that are going to be read during any further interval.
Extending this oracle, we moreover define the \textit{Oracle Modified (OM)} strategy that only saves the alive words that were modified in the current interval.
As for the MA scheme, we can consider that a complete snapshot of the system memory is stored in the NVM at the beginning and during any previous interval.
With the OM strategy, the data that will be read in the future are only saved if they were modified.
If a data has been saved in the previous intervals and remained unchanged, it is not added in the snapshot of the memory to be saved before the next power failure.
From now on, we will use Oracle to refer to the Oracle Modified when comparing with the other strategies. 

\subsection{Block-Based Strategies}
Both the \textit{Used Address} and \textit{Modified Address} strategies can be implemented with different degrees of granularity.
Tracking each individual word may require a very large memory to store the modified addresses, block-based strategy tries to trade-off between hardware cost and backup saving.
Instead of considering single word addresses, the addresses can be grouped in blocks of $N$ words and the scheme can be adapted to keep track of these blocks.
Therefore the \textit{Modified Block (MB)} strategy keeps track of the blocks that are modified during the interval. 
The backup size is given, for each interval, by the number of blocks that are accessed with one or more store operations.
In Freezer, the modified blocks are tracked using corresponding \textit{dirty bits}, which allows for the size of the associated tracking memory to be reduced by a factor equivalent to the block size.
MB with blocks of $N=1$ word corresponds to the MA strategy.

\section{Trace Analysis and Improvement in Backup Size}\label{sec:traces}
In order to validate our approach, we analyzed the memory access traces of several benchmarks from a subset of MiBench (see Table~\ref{tab:backup-size-vs-block-size} for a list of the benchmarks).
The benchmarks were run on a cycle accurate, bit accurate RISC-V model~\cite{rokicki:hal-02303453}, thus only two types of memory access are possible: load and store operations.
The traces report the information about each memory access during the program execution. 
In particular, each trace records  a timestamp (cycle count), the type of operation (ST or LD for store or load) and the address for every memory access.
Table \ref{tab:trace} shows an example of a memory access trace.
\begin{table}[htb]
    \caption{Example of memory access trace.}
    \centering
    \begin{tabular}{rlll}
        \hline
    interval & cycle &  op & addr \\
        \hline
        $i$ & ... & ... &     ...  \\
        $i$ & 90 &  ST &  0x38aaad4 \\
        $i$ & 97 &  LD &    0x2ba50 \\
        $i$ & 99 &  LD &    0x2b06c \\
        \hline
        $i+1$ & 104 &  LD &    0x2b954 \\
        $i+1$ & 109 &  LD &  0x38aaad4 \\
        $i+1$ & ... & ... &     ...  \\
        \vdots & \vdots & \vdots & \vdots \\
        $n$   & ... & ... &     ...  \\
        \hline
    \end{tabular}
    \label{tab:trace}
\end{table}
The occurrences of power failures are simulated by dividing an access trace in $n$ time intervals. 
Each interval $i$ is composed of a given number of clock cycles $N_{prog_i}$, equal to the active time $t_a$ of the interval $i$ divided by the processor clock period.
The cycle count reported in the trace is used to divide the execution of a benchmark in these $n$ intervals.
In the rest of the paper, for simplicity without loosing generality, we divide $t_{prog}$, the time needed for running the whole program without interruptions, in $n$ equal intervals of $N_{prog}$ cycles.

In the example reported on Table \ref{tab:trace}, the interruption is placed after cycle \textit{99}.
This means that the load happening in cycle \textit{104} is considered as being executed in the next interval ($i+1$).
This is a simple way to simulate a frequency of power failures every $N_{prog}$ cycles ($N_{prog}=100$ in this example).
In practice, for our simulations we considered longer intervals, ranging from $10^5$ to $10^7$ cycles.
As an example, considering a device running at 10 MHz, intervals of $10^6$ clock cycles would correspond to a frequency of interruptions due to power failures of 10 Hz.
In Section \ref{sec:impact-of-interval}, we present an analysis of the impact of the interval length and of the variability of intervals duration on the reduction of backup size. 

\begin{figure}[htb]
    \centering
    \includegraphics[width=.9\linewidth]{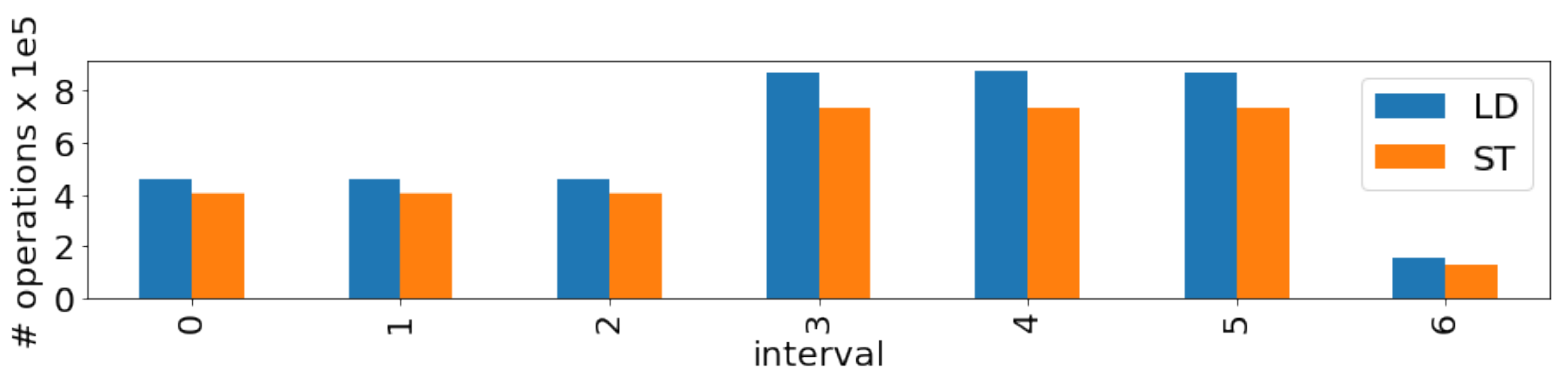}
    \caption{Number of LD and ST opserations per interval during the FFT benchmark execution with $N_{prog}=10^7$ clock cycles.}
    \label{fig:op}
\end{figure}

From these traces, the number of load and store operations per interval, as well as other memory access features, can be extracted.
As an example, Fig. \ref{fig:op} shows the number of LD and ST in each interval during the execution of the FFT benchmark, with intervals of $N_{prog}=10^7$ clock cycles. 
Considering the duration of the full execution of the FFT benchmark on the target processor, $n=7$ intervals can be simulated, ranging from interval $0$ to $6$ in the figure.
These traces provide relevant information about the memory access behavior of a given program. They will be used to compare the different backup strategies in Sections \ref{sec:Impact-of-block-size} and \ref{sec:results}.

\begin{figure}[ht]
    \centering
    \includegraphics[width=.9\linewidth]{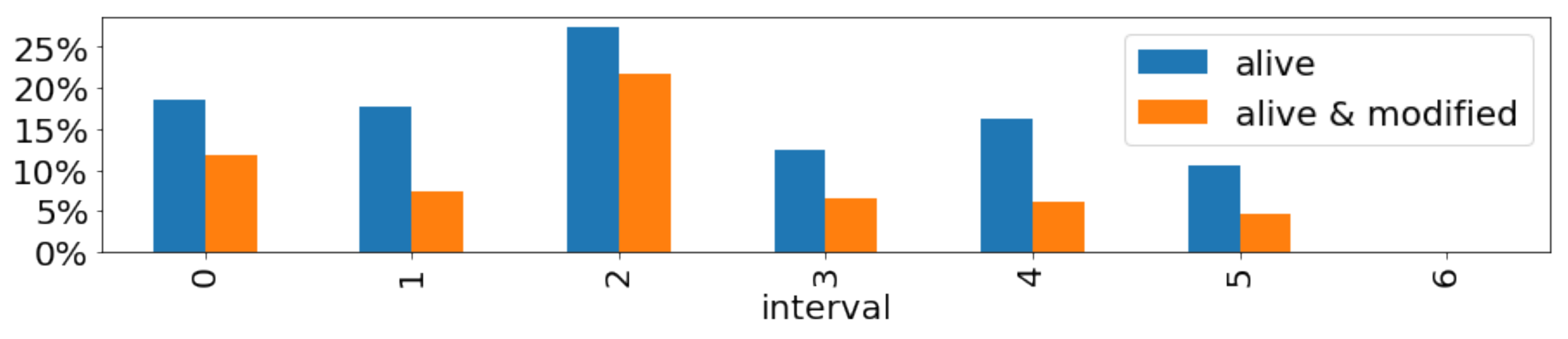}
    \caption{Percentage w.r.t. full memory space of ``alive'' and ``alive \& modified'' addresses per interval during the execution of the FFT benchmark with $N_{prog}=10^7$ clock cycles.}
    \label{fig:alive-fraction}
\end{figure}

Fig. \ref{fig:alive-fraction} shows the fraction of \textit{alive} and \textit{alive \& modified} addresses with respect to the total number of words addressed, for every interval of $10^7$ clock cycles for the FFT benchmark.
In the last interval no address is considered alive as the oracle knows that the program is going to terminate before the next power failure.
The figure also shows that, even with a relatively small benchmark, the number of words that really needs to be saved is less than a quarter of the total.
This motivates our work on the definition of new backup strategies to reduce the volume of data to be backed-up before a power failure.
However, as already mentioned, the OM cannot be implemented in a real system as it requires knowledge of the future.
It is however very useful for comparison as it gives the optimal lowest bound to the backup size.

\begin{figure}[htb]
    \centering
    \includegraphics[width=\linewidth]{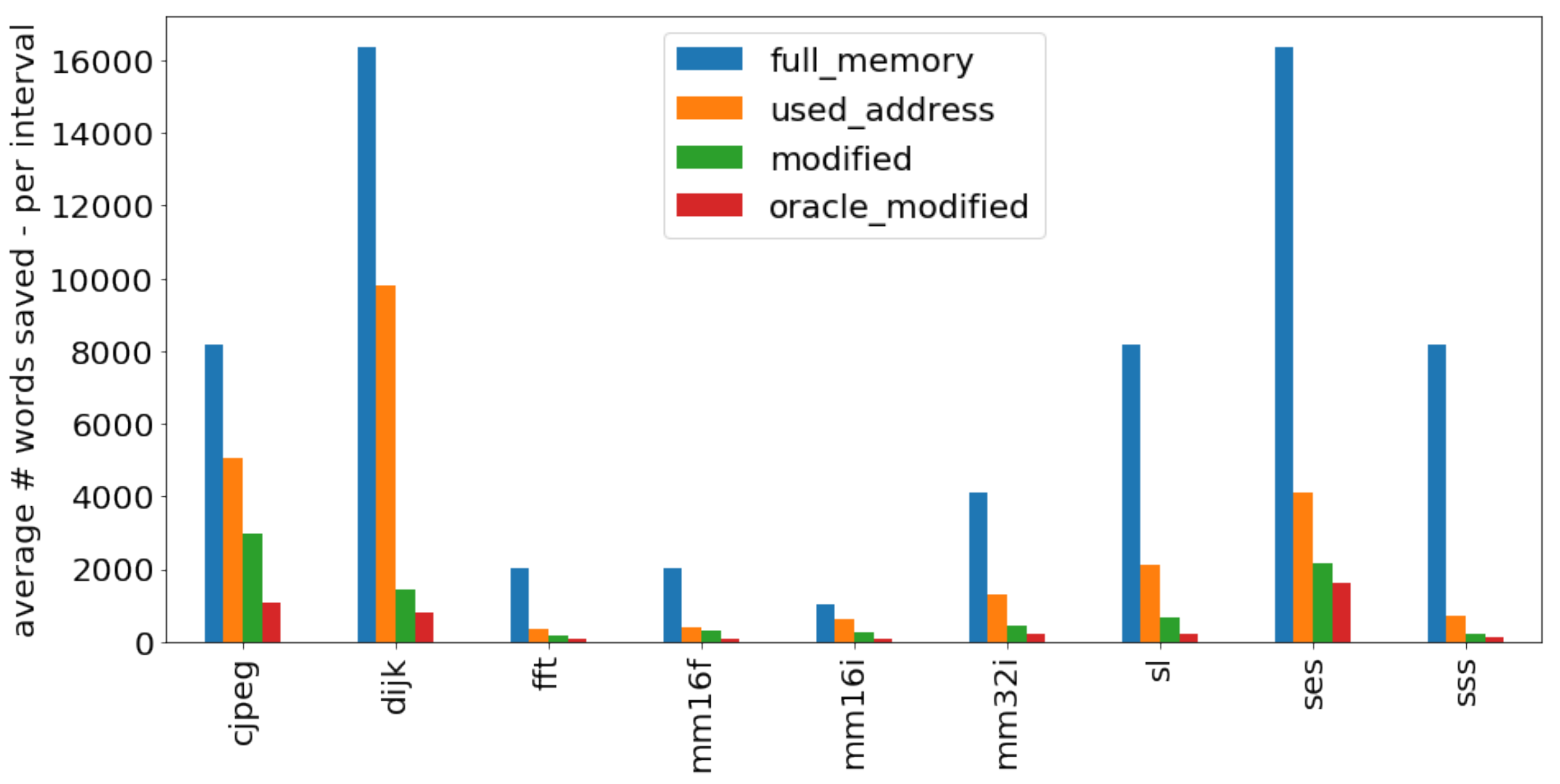}
    \caption{Average number of words saved per interval by the different backup strategies -- full-memory, used-address (UA), modified (MA), and oracle modified (OM) -- during the execution of different benchmarks, with $N_{prog}=10^6$ cycles.}
    \label{fig:ma-is-better}
\end{figure}
Fig. \ref{fig:ma-is-better} compares the average number of word saved per interval by the full-memory, UA, MA, and OM strategies for different benchmarks and with $N_{prog}=10^6$ clock cycles.
The figure shows the great potential of the proposed strategies w.r.t. state-of-the-art approaches.
Fig. \ref{fig:ma-is-better} also demonstrates that the MA strategy always outperforms the UA strategy in terms of number of saved words and it is the only technique that comes close to the performance of the oracle modified.
Therefore, only the MA strategy will be considered in the rest of the paper, as well as its extension to a block-based strategy presented in the following section.

\section{Freezer}\label{sec:freezer}

In this section, we present Freezer, a backup controller that implements the \textit{Modified Block} backup strategy, and study the impacts of the block size in the MB strategy.

\subsection{Freezer Architecture}
\begin{figure}[ht]
    \centering
    \includegraphics[width=.9\linewidth]{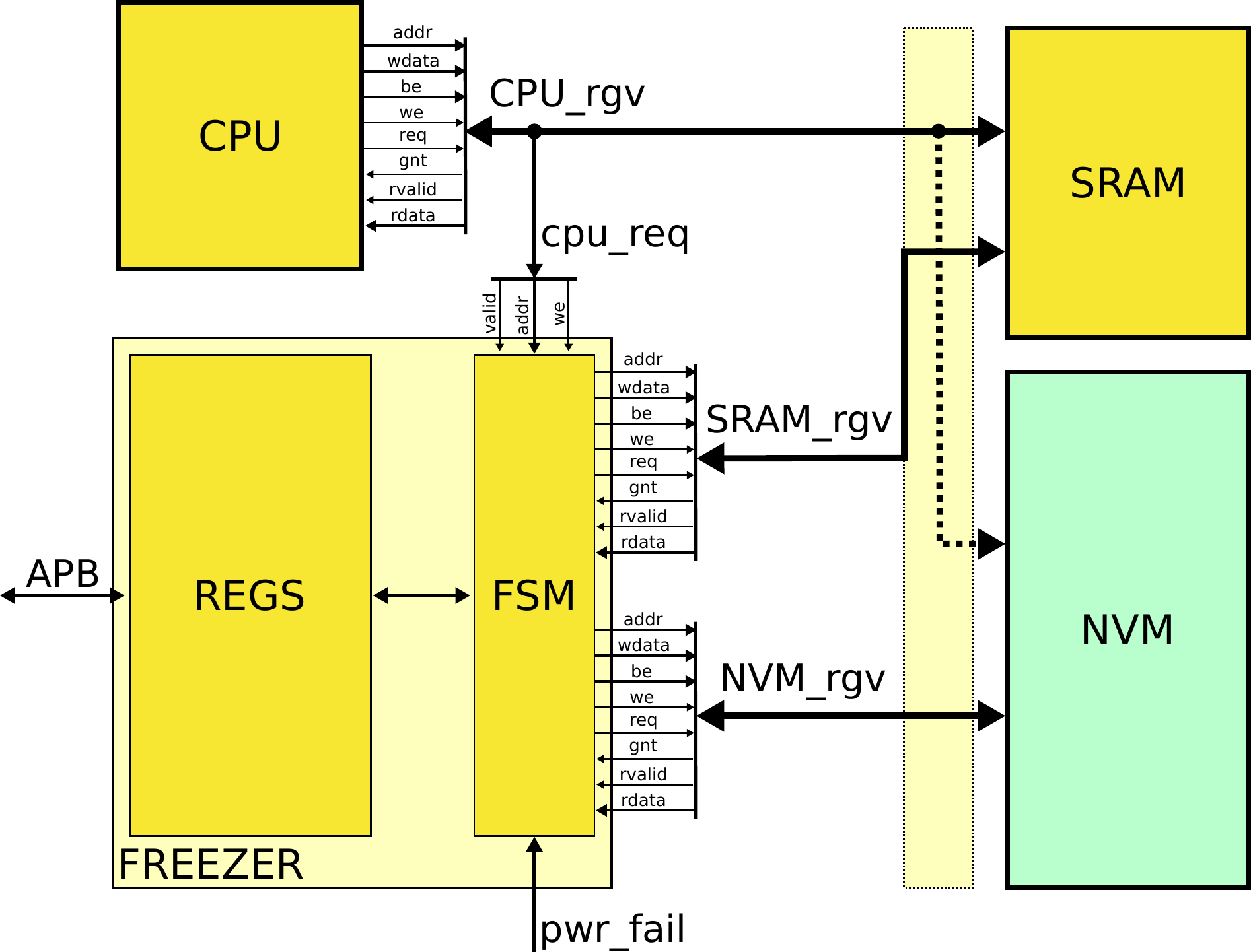}
    \caption{Freezer internal architecture}
    \label{fig:freezer-arch-int}
\end{figure}

Fig. \ref{fig:system-arch} shows the system-level view of the \textit{Freezer} architecture.
The system is composed of four major components: the CPU, the SRAM used as a main memory, the NVM used for the backup, and the backup controller (Freezer).
Freezer is itself composed of two main blocks: a controller implemented as a finite-state machine (FSM) for sequencing the operations and a small memory containing the dirty bits used to keep track of the blocks that need to be saved, as shown in Fig.~\ref{fig:freezer-arch-int}.
The Freezer controller is a stand-alone component, that does not need to be tightly coupled with the memories or with the core.
It uses two handshake interfaces for the SRAM and NVM requests, allowing to tolerate variable access latency. 
Freezer can be directly connected to the control, address and data signals of both SRAM and NVM, using these handshake interfaces.
Alternatively the SRAM and NVM interfaces can be arbitrated and share a single master port on the system bus.
Moreover, Freezer is also connected to the request signals of the CPU to the SRAM, this allows Freezer to (i) spy the address of the SRAM accesses by the processor and (ii) manage the backup-to and restore-from-NVM phases in place of the processor.
SRAM and NVM do not need to have two ports, CPU and Freezer accesses can be easily arbitrated as they never access the memory at the same time.

At run-time, Freezer checks the address of the store operations in the SRAM to dynamically keep track of the blocks that are modified.
When a power failure arises, the CPU is halted and the controller starts transferring the modified blocks into the non-volatile memory. 
The words within a block are then stored sequentially in the NVM.
The controller uses the information collected during the active time to determine which blocks to save.
When performing this task, Freezer has access to both the SRAM and the NVM memory. 


Algorithm \ref{algo:backup-controller} describes the behavior of the backup controller during the execution, backup, and restore phases.
During execution, Freezer implements the \textit{Modified Block} backup strategy. 
\begin{algorithm}[tb]
\SetKwIF{If}{ElseIf}{Else}{if}{:}{elif}{else:}{}%
\SetKwFor{For}{for}{\string:}{}%
\SetStartEndCondition{ }{}{}%
\AlgoDontDisplayBlockMarkers\SetAlgoNoEnd\SetAlgoNoLine

\caption{Freezer backup controller algorithm}
\label{algo:backup-controller}

\KwIn{cpu\_addr address generated by the CPU}
\KwIn{is\_store = 1 if the operation is a store}
\KwIn{op\_valid = 1 if the operation is valid}
\KwIn{pwr\_fail = 1 if power failure is detected}
\KwIn{restore = 1 if resume after a power failure} 
\BlankLine
\KwData{\textit{to\_backup} flag memory of 1-bit per block}
\BlankLine
\eIf{restore}{
    \For{i $\gets 0$ \textbf{to} $SRAM\_SIZE-1$}{
        sram[i] $\gets nvm[i]$\;
    }
}
{
    \eIf{\textbf{not} pwr\_fail}{
        \If{is\_store \textbf{and} op\_valid}{
        block $\gets cpu\_addr \gg log2(BLOCK\_SIZE)$\;
        to\_backup[block] $\gets 1$\;
        }
    }{
        \For{$b \gets 0$ \textbf{to} $BLOCK\_NUM-1$}{
            \If{to\_backup[b]}{
                \For{$a \gets 0$ \textbf{to} BLOCK\_SIZE-1}{
                    addr $\gets (b \ll log2(BLOCK\_SIZE)) \| a$\;
                    nvm[addr] $\gets$ sram[addr]\;
                }
            }
        }
    }
}
\end{algorithm}
During execution, when there is no power failure (not \textit{pwr\_fail}) and there is a valid store operation, the controller records the blocks that are modified in a table (\textit{to\_backup}) implemented in a small memory, or in a register bank.
When the \textit{pwr\_fail} condition is true, it enters in the backup phase and in a loop where, for each block, the \textit{to\_backup} memory is checked.
If the block has to be saved, then a loop for every address of the block is executed, where a word is read from the SRAM and written in the NVM.
This last loop can easily be pipe-lined such that an NVM write in an address can be executed in the same cycle with an SRAM read in the successive address.
The same holds true also for the restore phase, that simply moves back the data from the NVM to the SRAM.
In this way, the backup controller is able to back up and restore one word every clock cycle.
This should also lead to an additional speed-up, when compared with software-based backup loops executed on low-end micro-controllers, as in the case of \cite{balsamo_hibernus:_2015}.

In the hardware implementation, the process of checking the dirty bits can also be optimised.
As an example, the scan of the \textit{to\_backup} memory to find the next dirty block can happen in parallel to the backup of the current block, which is a relatively long operation.
Moreover, the \textit{to\_backup} memory can be organised as a matrix of dirty bits and the controller can check an entire row of dirty bits in parallel.
This means that the \textit{to\_backup} memory can be scanned row by row.
The sparsity of the dirty bits can also be exploited: skipping rows that have only clear bits (all zeros).

With these and other optimisations, the throughput of the backup operation can be sustained with little to no dead cycles.
However these low level optimisations are outside the scope of this work and will not be investigated further.

\subsection{Area and Power Results}
\label{sec:areapowefreezer}

As our algorithm is relatively simple, the controller itself introduces small area and power overheads. 
The major contribution in the area and power overheads is given by the \textit{to\_backup} dirty-bit memory, used to keep track of the blocks that have to be saved. 
Table \ref{tab:flag-mem} shows the number of bits and an estimation of the area of the \textit{to\_backup} memory for different block sizes, considering a 32KB SRAM.
\begin{table*}[ht]
    \centering
    \caption{Number of bits and area estimation of the \textit{to\_backup} memory, implemented with standard cells in 28 nm FDSOI.}
    \begin{tabular}{c|c|c|c|c|c|c}
        \hline
        block size (32bit words) & 2        &  4      & 8        & 16       & 32      & 64      \\\hline
        \# bits                  & 4196     & 2048    & 1024     & 512      & 256     & 128     \\ 
        area       $[\mu m^2]$   & 10838.11 & 5452.02 & 2748.94  & 1436.81  & 730.64  & 386.95  \\
        \hline
    \end{tabular}
    \label{tab:flag-mem}
\end{table*}
For these results, the \textit{to\_backup} memory is synthesized with standard cells in a 28nm FDSOI technology using Synopsys Design Compiler (DC).
Even when considering a fine granularity for the block size, the dirty-bit memory is small compared to the total size of the SRAM memory.
As an example, for a block size of $8$ words, the required 1024-bit memory is $256\times$ smaller than the main SRAM memory.
Moreover, by tuning the block size with larger blocks, the \textit{to\_backup} memory can be stored in a register file with a small increase in the backup size.

A non optimized version of the controller was synthesized from a C++ specification using Mentor Graphics CatapultHLS and Synopsys (DC) with the same 28nm FDSOI technology at $0.7V$.
In this configuration, Freezer's controller achieves a dynamic power of $P_{active} = 6.8 \mu W$ and a leakage power of around $P_{leak} = 40 nW$ at 25\textdegree.
With the same technology, we estimate a leakage of roughly $600 nW$ for a register-based \textit{to\_backup} memory of $1024$ bits.
These synthesis results will be exploited in Section \ref{sec:area} to estimate the energy of a system implementing Freezer.

\subsection{Impact of Block Size} \label{sec:Impact-of-block-size}
In this section, we study the impact of the block size on the size of the backup provided by the MB strategy.
The size of the \textit{to\_backup} memory depends on two parameters: the number of 32-bit words in each block, which determines the granularity of the backup strategy, and the total size of the SRAM.
Therefore, it is possible to trade off an increase in backup size with a smaller area overhead of the \textit{to\_backup} memory.
In Table \ref{tab:backup-size-vs-block-size}, the backup size across a set of benchmarks is reported for different configurations of block granularity.
The backup size is averaged on all intervals and normalized with respect to a block of one word (MA strategy). The interval is set to $N_{prog}~=~10^6$ clock cycles.
\begin{table}[htb]
    \centering
    \caption{Backup size relative to blocks of one 32-bit word (MA approach) for different benchmarks. $N_{prog}=10^6$ cycles. The table also reports the average on all benchmarks.}
    \begin{tabular}{l|rrrrrr}
        \hline
        block size  $N$             &  2     &  4     &  8     &  16    &  32    &  64 \\
        \hline                 
        susan\_smooth\_small    (sss) &  1.01  &  1.04  &  1.09  &  1.17  &  1.32  &  1.62 \\
        susan\_edge\_small      (ses) &  1.03  &  1.10  &  1.22  &  1.35  &  1.54  &  1.68 \\
        matmul16\_float       (mm16f) &  1.11  &  1.24  &  1.47  &  1.77  &  2.23  &  2.54 \\
        qsort                 (qsort) &  1.01  &  1.03  &  1.07  &  1.28  &  1.58  &  1.70 \\
        fft                     (fft) &  1.10  &  1.22  &  1.44  &  1.79  &  2.45  &  3.22 \\
        matmul32\_int         (mm32i) &  1.03  &  1.08  &  1.17  &  1.28  &  1.47  &  1.75 \\
        str\_search             (str) &  1.02  &  1.06  &  1.12  &  1.17  &  1.28  &  1.55 \\
        cjpeg                 (cjpeg) &  1.01  &  1.03  &  1.06  &  1.10  &  1.18  &  1.32 \\
        dijkstra               (dijk) &  1.06  &  1.18  &  1.31  &  1.36  &  1.45  &  1.55 \\
        matmul16\_int         (mm16i) &  1.07  &  1.19  &  1.38  &  1.69  &  2.16  &  2.76 \\
        susan\_edge\_large      (sel) &  1.08  &  1.17  &  1.30  &  1.46  &  1.59  &  1.72 \\
        \hline                
        average                 (avg) &  1.05  &  1.12  &  1.24  &  1.40  &  1.66  &  1.95 \\
        \hline
    \end{tabular}
    \label{tab:backup-size-vs-block-size}
\end{table}
Increasing the block size has obviously an impact on the performance of the MB strategy, i.e., the average size of the backup required at the end of each interval.
However, MB with a relative small size of block (up to $N=8$) only increases the backup size by 24\% in average, while this block-based strategy can decrease the size of the \textit{to\_backup} memory by a factor of $8\times$.
More comparisons and impact of the interval size are reported in the next section.

\section{Results}\label{sec:results}
In this section, the details of the experimental setup are explained and the results regarding the backup size (Sec.~\ref{sec:res:backupsize}) and backup time {(Sec.~\ref{sec:res:backuptime})} are reported.
The impact of the interval size is discussed in Section \ref{sec:impact-of-interval}. For all the other results provided in this section, the interval is set to $N_{prog}=10^6$ clock cycles.
Moreover, a discussion about power, energy, and area of our approach is presented in Sections \ref{sec:mem-energy-cmp} and \ref{sec:area}, 
while considerations on the impact of leakage are presented in Section \ref{sec:leakage}.

\subsection{Backup Size}\label{sec:res:backupsize}
\begin{figure}[ht]
\centering
\includegraphics[width=.95\linewidth]{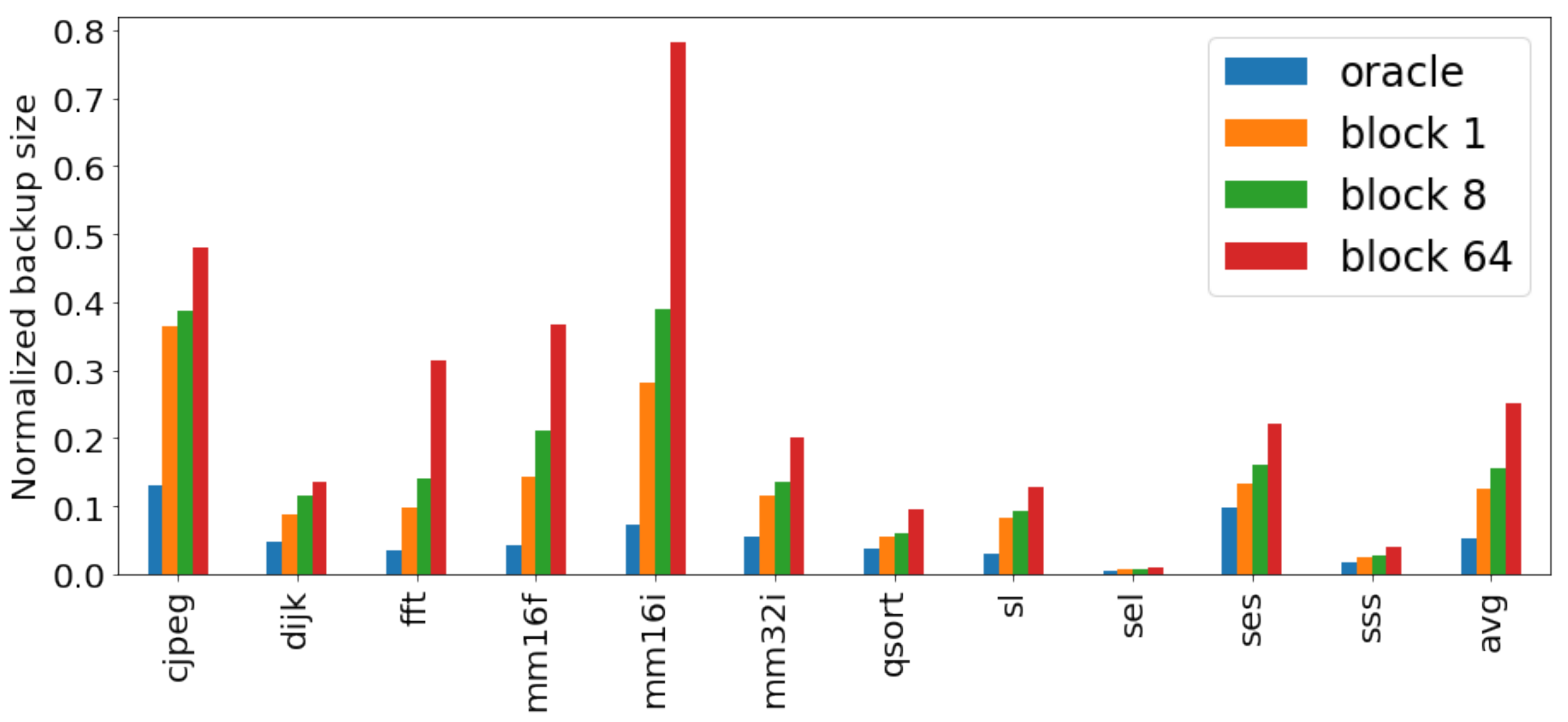}
\caption{Backup size normalized w.r.t. the program memory size (improved Hibernus strategy) of Freezer implementing MB strategy with blocks of 1, 8, and 64 words. Lower bound in backup size of the oracle-modified is also reported.}
\label{fig:backup-size-plt}
\end{figure}

For every interval, the backup size is computed considering the different approaches described in Section \ref{sec:backup-model}.
Fig. \ref{fig:backup-size-plt} shows the backup size reported for every benchmark and for blocks of 1, 8, and 64 32-bit words.
Blocks of size equal to one word corresponds to the MA strategy.
The \textit{OM} strategy is also reported to provide the optimal (non reachable) value. 
The backup size is averaged on all intervals and normalized against the improved Hibernus~\cite{balsamo_hibernus:_2015} approach, which saves the full memory used by the program in pages of 512 bytes.

As it can be seen from Fig. \ref{fig:backup-size-plt}, our approach greatly reduces the average backup size per interval, reaching an $87.7\%$ (more than $8\times$) reduction in average, with only a $7.5\%$ distance from the \emph{oracle modified}, when configured with a granularity of $8$ words per block.
This reduction in backup size can be directly converted into an energy saving in number of write to the NVM during the backup phase.

\subsection{Impact of Interval Size}\label{sec:impact-of-interval}
\begin{figure}[htb]
\centering\includegraphics[width=.95\linewidth]{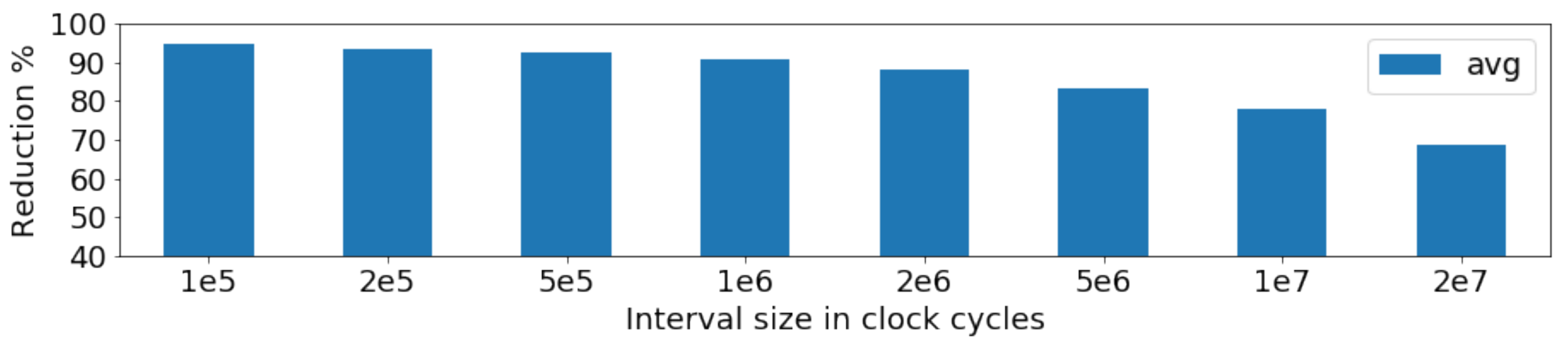}
\caption{Average backup size reduction with different interval size $N_{prog}$ expressed in number of clock cycles.}
\label{fig:backup-iv-size}
\end{figure}

On a system powered with intermittent ambient energy, the time length of the intervals is mostly determined by the energy source and by the energy budget of the platform.
If the energy source is relatively stable, the length of the power cycle increases, and so does the amount of computation that the processor manages to complete during one interval.
This means that more memory accesses will be performed, thus we can expect the average size of a backup to increase.
However, this also depends on the spatial locality of the application, and considering wider blocks could be beneficial for less intermittent sources.
When the length of the power cycles decreases, the processor is interrupted more frequently, and the number of memory accesses is reduced.
Therefore, the average backup size is further decreased.
Fig. \ref{fig:backup-iv-size} shows the average reduction in backup size, across all benchmarks, for different lengths of the power cycles (interval size $N_{prog}$ expressed in number of clock cycles), considering blocks of 8 words.
As it can be seen, the backup size reduction is greater than $70\%$ for all the interval lengths.
Moreover, with shorter intervals, the reduction becomes greater than $90\%$.
It must be noted also that, when the length of the interval is increased above 20 million clock cycles, the majority of the programs are able to run to completion before the first power failure occurs.

\begin{figure}[ht]
    \centering
    \begin{subfigure}{.48\textwidth}
    \centering
    \includegraphics[width=\linewidth]{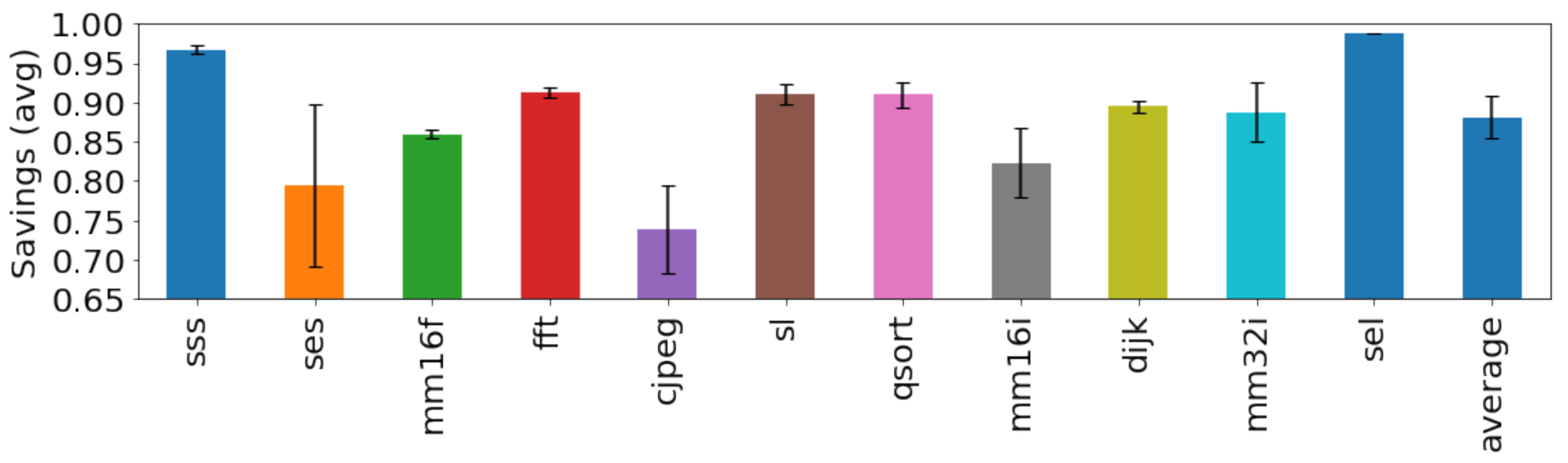}
    \caption{Failure rate $10^{-6}$.}
    \label{fig:random-iv-1e6}
    \end{subfigure}
    \begin{subfigure}{.48\textwidth}
    \centering
    \includegraphics[width=\linewidth]{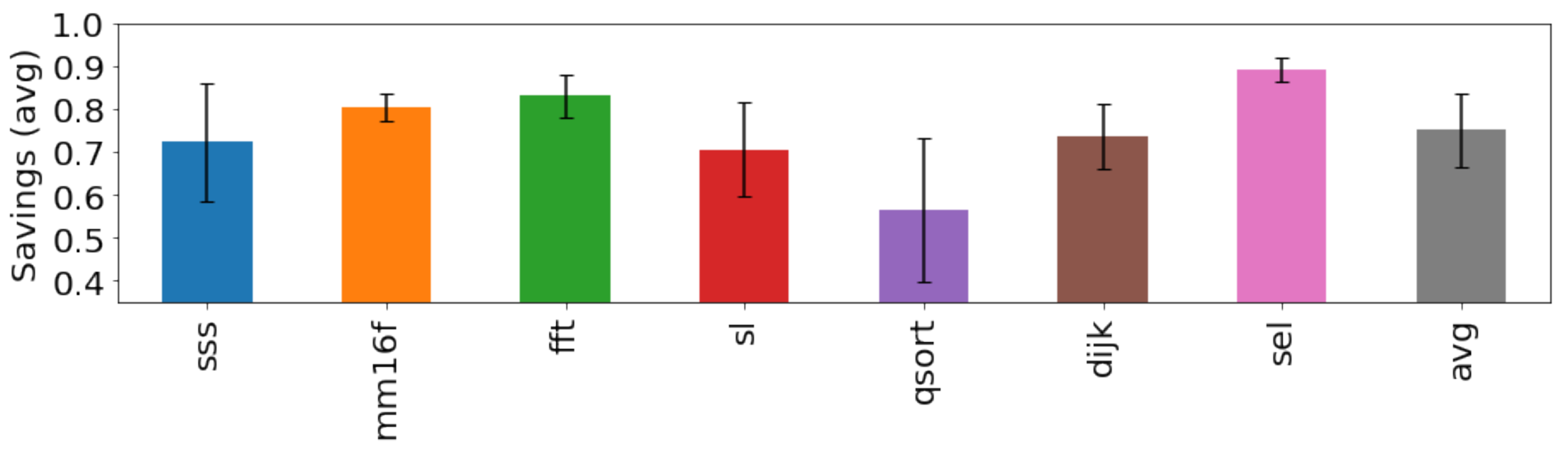}
    \caption{Failure rate $10^{-7}$.}
    \label{fig:random-iv-1e7}
    \end{subfigure}
    \caption{Average savings and std. deviation for 100 executions with power failures distributed following the binomial law with failure rates of $10^{-6}$ (a) and $10^{-7}$ (b).}
    \label{fig:random-iv}
\end{figure}

Due to the unpredictability of the energy source, an intermittently-powered system might also experience a wide variation between the time length of successive intervals.
To better capture this behaviour, we model the occurrence of a power failure as a random variable distributed according to a binomial law.
Power failure events in this model are considered independent of one another. 
At each clock cycle, there is a certain probability to incur in a power failure.
For this experiment, we considered two values of one power failure every $10^6$ cycles and one power failure every $10^7$ clock cycles.
Figures \ref{fig:random-iv-1e6} and \ref{fig:random-iv-1e7} show, for each benchmark, the average savings with relative standard deviation computed for 100 executions, considering blocks of 8 words.
Our proposed method is robust to variability in the size of the intervals and it is able to achieve more than $83\%$ and $88\%$ savings on average when the failure rates are respectively $10^{-7}$ and $10^{-6}$.

\subsection{Backup Time}\label{sec:res:backuptime}
The reduction in the backup size comes with a relative reduction in the save time.
On top of that, thanks to the hardware accelerated backup process, our solution provides an additional improvement in terms of backup time.
In particular, the backup process is managed directly by Freezer and can be further pipelined, so that each word can be saved in one clock cycle.
Of course, the speed of this process is limited by the cycle-time of the slowest NVM memory.
As our approach does not rely on any specific NVM technology, we considered the numbers reported in \cite{balsamo_hibernus:_2015} for our comparison.
In particular, we considered a clock frequency of 24 MHz for the normal operation using SRAM, and a clock cycle period of 125 ns (8 MHz) for the FeRAM.

\begin{table}[htb]
\centering
\caption{Percentage reduction of backup time w.r.t. improved Hibernus (higher is better). Columns b\_$N$ provides results with our strategy using blocks of $N$ words. Oracle modified and NVP \cite{liu_4.7_2016} are also provided for comparison.}
\begin{tabular}{lccccc}
\hline
{}                   &  b\_1   &  b\_8   &  b\_64  &  oracle &  NVP \cite{liu_4.7_2016}\\
\hline
susan\_smooth\_small &  99.80  &  99.78  &  99.68  &  99.87  &  39.25 \\
susan\_edge\_small   &  98.93  &  98.69  &  98.20  &  99.20  &  42.58 \\
matmul16\_float      &  99.32  &  99.00  &  98.27  &  99.79  &  39.08 \\
qsort                &  99.38  &  99.34  &  98.95  &  99.58  &  45.96 \\
fft                  &  99.58  &  99.39  &  98.64  &  99.85  &  39.64 \\
matmul32\_int        &  99.35  &  99.23  &  98.86  &  99.69  &  40.44 \\
str\_search          &  99.49  &  99.43  &  99.21  &  99.81  &  43.03 \\
cjpeg                &  98.10  &  97.98  &  97.48  &  99.32  &  38.97 \\
dijkstra             &  99.34  &  99.13  &  98.98  &  99.63  &  43.32 \\
matmul16\_int        &  99.23  &  98.94  &  97.88  &  99.80  &  29.94 \\
susan\_edge\_large   &  99.93  &  99.91  &  99.89  &  99.95  &  45.78 \\
\hline                                                       
average              &  99.31  &  99.17  &  98.73  &  99.68  &  40.73 \\
\hline
\end{tabular}
\label{tab:backup-time}
\end{table}

Table \ref{tab:backup-time} reports the improvement in backup time compared with a modified implementation of Hibernus that only saves the memory used by the program (in pages of 512 KB).
Columns b\_$N$ provides results with our strategy using blocks of $N$ 32-bit words.
For the column related to non-volatile processor (NVP), we considered the backup time reported in~\cite{liu_4.7_2016} of 1.02 ms for 4KB, and scaled it for the memory size of our benchmarks, grouping the addresses in pages of 1 KB.
With this configuration, our approach gives a two orders of magnitude improvement in backup time when compared to the software-based approach that saves the whole program memory.
Moreover, Freezer provides a significant advantage also when compared with a fully non-volatile processor as \cite{liu_4.7_2016}, which only provides an improvement of 40\% when compared to the software-based approach. 

This improvement in the backup time is also going to affect positively the total execution time, as expressed in Eq. \ref{eq:time}.
We considered a 24~MHz frequency for the volatile operations and an 8~MHz frequency for the FeRAM accesses.
The active time is set to $N_{prog}~=~10^7$ clock cycles at 24~MHz.
We assumed an average off time equal to the active time. 
As reported in Table~\ref{tab:exec-time}, our strategy achieves a 32\% average decrease of the total execution time when compared with improved Hibernus.
We also compared Freezer against approaches like QuickRecall \cite{jayakumar_quickrecall:_2015} that runs the programs only using the NVM. 
In this case, the save and restore time are roughly zero (only the registers need to be saved), but the frequency of the core is limited to the frequency of the FeRAM.
As a consequence, in most cases, the QuickRecall approach leads to longer execution time than Hibernus, whereas our solution performs always better and is very close to the Oracle.

\begin{table}[htbp]
\centering
\caption{Percentage reduction of execution time w.r.t. improved Hibernus (higher is better). Columns b\_$N$ provides results with our strategy using blocks of $N$ words. Oracle and NVM-only solution of \cite{jayakumar_quickrecall:_2015} are also provided for comparison.}
\begin{tabular}{lcccc}
\hline
{} &  b\_1 &  b\_8 &  oracle &   NVM only \cite{jayakumar_quickrecall:_2015}\\
\hline
susan\_smooth\_small &     20.75 &     20.75 &   20.76 & -17.59 \\
susan\_edge\_small   &     33.35 &     33.31 &   33.41 &   2.35 \\
matmul16\_float     &      9.90 &      9.88 &    9.93 & -34.50 \\
qsort              &     88.79 &     88.77 &   88.90 &  89.00 \\
fft                &     10.68 &     10.67 &   10.70 & -33.30 \\
matmul32\_int       &     15.32 &     15.31 &   15.35 & -26.01 \\
str\_search         &     24.90 &     24.89 &   24.95 & -11.05 \\
cjpeg              &     27.93 &     27.91 &   28.13 &  -5.94 \\
dijkstra           &     35.21 &     35.17 &   35.27 &   5.14 \\
matmul16\_int       &      8.72 &      8.71 &    8.75 & -36.34 \\
susan\_edge\_large   &     83.49 &     83.48 &   83.50 &  80.27 \\
\hline
average            &     32.64 &        32.62 &   32.69 &   1.09 \\
\hline
\end{tabular}
\label{tab:exec-time}
\end{table}

\subsection{Energy Comparison with other Memory Models}
\label{sec:mem-energy-cmp}
\begin{table*}[htb]
    \centering
    \caption{Energy and leakage power parameters used for memory access cost simulation. Read/write energy is reported in pJ per 32-bit word access.}
    \begin{tabular}{l|llll|llll|llll}
        \hline
                             & \multicolumn{4}{l|}{SRAM}     & \multicolumn{4}{l|}{STT}          & \multicolumn{4}{l}{RRAM}         \\ \hline
        Size  {[}KB{]}       & 4     & 16    & 32    & 64    & 4      & 16     & 32     & 64     & 4      & 16     & 32     & 64     \\ \hline
        Read  {[}pJ{]}       & 0.219 & 0.703 & 1.664 & 2.50  & 7.754  & 7.889  & 8.426  & 8.692  & 5.101  & 5.477  & 6.004  & 6.667  \\ \hline
        Write {[}pJ{]}       & 0.111 & 0.215 & 1.175 & 1.388 & 20.244 & 20.614 & 20.873 & 21.416 & 21.349 & 27.449 & 24.176 & 28.575 \\ \hline
        Leakage {[}$\mu$W{]} & 0.78  & 2.16  & 3.58  & 7.16  & \multicolumn{8}{l}{~} \\ \cline{1-5}
    \end{tabular}
    \label{tab:mem-energy-values}
\end{table*}
%
\begin{table}[htb]
    \centering
    \caption{Cache Miss and Hit dynamic energy in pJ per 32-bit word access}
    \begin{tabular}{c|c|c}
        \hline
        Size [KB] & Hit Energy [pJ] & Write Energy [pJ] \\
                  & $E_{cache/r}$   & $E_{cache/w}$     \\
        \hline
         2        & 5.43            & 4.5               \\
         4        & 6.15            & 4.96              \\
         8        & 10.13           & 9.42              \\
         16       & 13.45           & 12.74             \\
         \hline
    \end{tabular}
    \label{tab:cache-dynamic-energy}
\end{table}

\begin{figure}[htbp]
    \centering
    \begin{subfigure}{.49\textwidth}
    \includegraphics[width=.95\linewidth]{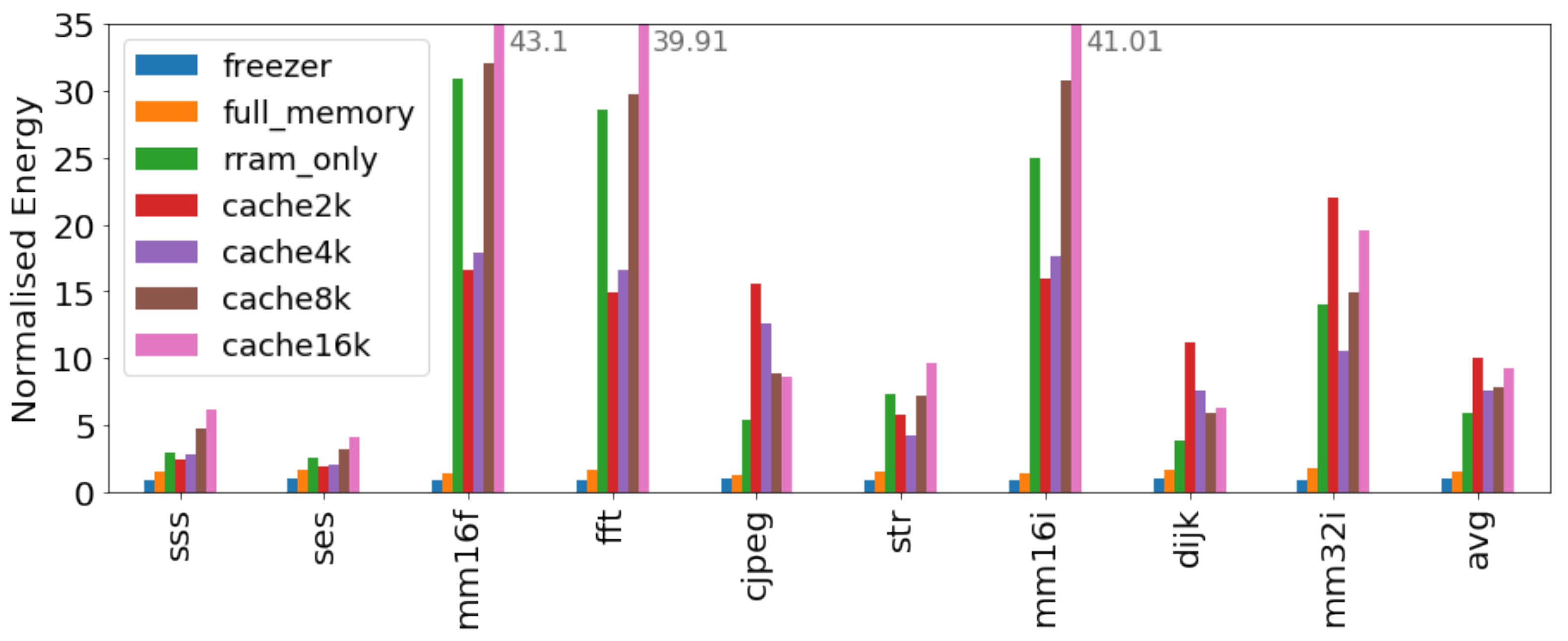}
    \caption{RRAM}
    \label{fig:rram}
    \end{subfigure}
    \begin{subfigure}{.49\textwidth}
    \includegraphics[width=.95\linewidth]{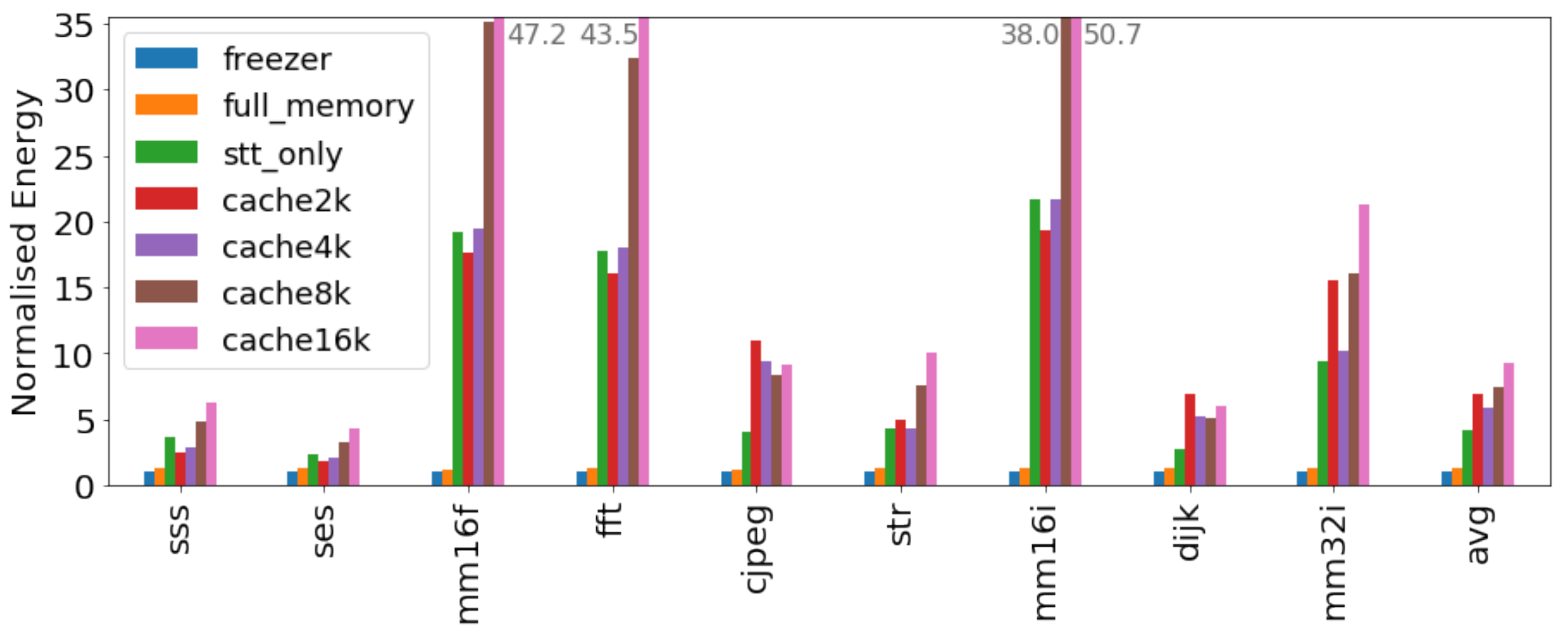}
    \caption{STT}
    \label{fig:stt}
    \end{subfigure}
    \caption{Relative dynamic energy of memory accesses, normalized w.r.t. Freezer, using RRAM (a) and STT (b) as NVMs for backup.}
    \label{fig:energy-cost}
\end{figure}

We use Eqs. (\ref{eq:energy-sram+NVM}), (\ref{eq:energy-NVM}), and (\ref{eq:energy-cache}) to compare the dynamic memory access energy of the different system configurations.
Figures~\ref{fig:rram} and~\ref{fig:stt} show these dynamic energies normalised w.r.t. the system using Freezer, with RRAM and STT, respectively.
For the cache+NVM architecture, four different cache sizes of 2KB, 4KB, 8KB and 16KB are reported. 
The three caches are all 4-way set associative with lines of 8 words (256 bits), which is representative of this type of device.
We considered blocks of 8 words also for the system using Freezer.
The read and write dynamic energies per 32-bit word for the memories used in these comparison are reported in Table \ref{tab:mem-energy-values}, and were obtained using NVSim \cite{dong_nvsim:2012}.
Table \ref{tab:cache-dynamic-energy} reports the Hit and Write dynamic energy for the different cache sizes, obtained with NVSim. 
Miss energies were in all cases equal to Hit energies.
As it can be seen from Fig.~\ref{fig:energy-cost}, our proposed approach provides a significant reduction in the energy due to memory accesses when compared with all the other methods.
The memory access energy for a full-memory backup strategy is in average $1.26\times$ of that required by Freezer when using STT and $1.65\times$ when considering RRAM.

Being based on the same SRAM+NVM architecture, Freezer and full-memory backup strategies require the same absolute amount of energy for the execution of the program, i.e., the energy required for executing load and store operations is the same for the same benchmark.
Moreover as the two strategies rely on a full memory restore after a power failure, they spend the same amount of energy for the restore memory accesses, for the same benchmark.
Tables \ref{tab:energy-percent-stt-freezer} and \ref{tab:energy-percet-stt-full-mem} show the energy decomposition, across all benchmarks, for Freezer and full-memory strategies when using STT NVM.
The two tables show the clear advantage that Freezer brings in terms of backup energy, reducing its weight from an average $23.25\%$ to an average $3.44\%$ of the total memory access energy.
\begin{table}[htb]
    \caption{Memory access energy percentage decomposition for Freezer using STT}
    \centering
    \begin{tabular}{lrrrr}
    \hline
    Trace & backup & restore & prog. loads &  prog. stores \\
    \hline
    sss   &   0.74 &   22.60 &       74.86 &   1.80 \\
    ses   &   5.97 &   29.30 &       59.23 &   5.49 \\
    mm16f &   5.53 &   21.89 &       49.88 &  22.70 \\
    fft   &   4.72 &   28.28 &       46.12 &  20.88 \\
    cjpeg &   7.53 &   16.30 &       57.64 &  18.53 \\
    str   &   2.62 &   23.86 &       43.65 &  29.87 \\
    mm16i &  15.13 &   33.12 &       40.55 &  11.21 \\
    dijk  &   3.45 &   23.56 &       61.34 &  11.64 \\
    mm32i &   4.46 &   26.86 &       60.42 &   8.26 \\
    \hline
    avg   &   3.44 &   23.52 &       61.18 &  11.86 \\
    \hline
    \end{tabular}
    \label{tab:energy-percent-stt-freezer}
\end{table}
\begin{table}[htb]
    \caption{Memory access energy percentage decomposition for full-memory backup using STT}
    \centering
    \begin{tabular}{lrrrr}
    \hline
    Trace & backup & restore & prog. loads &  prog. stores \\
    \hline
    sss   &  21.39 &   17.90 &  59.29      &  1.43 \\
    ses   &  28.29 &   22.35 &  45.18      &  4.19 \\
    mm16f &  21.66 &   18.15 &  41.36      & 18.82 \\
    fft   &  26.15 &   21.92 &  35.75      & 16.18 \\
    cjpeg &  17.40 &   14.56 &  51.49      & 16.55 \\
    str   &  22.65 &   18.95 &  34.67      & 23.72 \\
    mm16i &  31.33 &   26.80 &  32.81      &  9.07 \\
    dijk  &  23.60 &   18.65 &  48.54      &  9.21 \\
    mm32i &  25.77 &   20.87 &  46.94      &  6.42 \\
    \hline
    avg   &  23.25 &   18.69 &  48.63      &  9.42 \\
    \hline
    \end{tabular}
    \label{tab:energy-percet-stt-full-mem}
\end{table}
Figures \ref{fig:rram} and \ref{fig:stt} also show that, due to the higher read and write dynamic energies, using the NVM as the main memory is often detrimental even when compared with full-memory backup systems.
Moreover when compared to Freezer, NVM-only systems require in average $6.19\times$ and $4.22\times$ more energy for RRAM and STT respectively. 

As described in Section \ref{sec:res:energy}, the cache+NVM system uses the write-back policy and flushes the dirty lines in the NVM when a power failure arises. 
Thus the cache+NVM system shows a behaviour that is similar to the one of Freezer during power failures, but with higher energy per operation.
There are however some major differences between a system that implements Freezer and a system with a write-back cache and a NVM main memory. 
First of all, Freezer is meant to be simple to reduce the energy overhead of tracking the modified blocks.
Moreover, Freezer is able to track the full main memory and only needs to write on the NVM before a power failure happens.
A write-back cache on the other hand might perform additional writes on the NVM at run-time.
In fact if the access causes a conflict, the cache will evict the conflicting line thus causing additional NVM writes. 
These additional writes may reduce the lifetime of the NVM due to the limited endurance of these type of memories.

When it comes to cache+NVM based systems, the size that in average provides the smallest energy is 4KB, with 2KB and 8KB caches performing better in some benchmarks.
Access to smaller caches requires less energy, as shown in Table \ref{tab:cache-dynamic-energy}, but they might incur in the high cost of additional NVMs read and writes due to a larger number of misses and evictions. 
The increased number of writes to the NVM could also cause problems of endurance because of wear-out, that might prevent this solution to be applied for long-lasting operations.
A larger cache can reduce the number of accesses to the NVM, up to the point where the cache is so large that it is able to buffer the full application.
In these case, it is possible to obtain a number of writes to the NVM which is close to what Freezer achieves.
However, this comes at the cost of having a large cache that is complex and energy hungry.
Moreover, it is unusual to see a cache used in small low-power edge devices, where the system memory is embedded on chip and seldom exceeds 64KB.
To summarize, for our set of benchmarks, the energy required by a 4KB cache + STT system is $5.9 \times $  w.r.t. Freezer, whereas the larger 16KB cache requires in average $9.3 \times$ more energy than Freezer.
 
\subsection{Impact of Leakage Power}

\label{sec:leakage}
For a fair comparison, it is also important to study the impact of leakage power of the {SRAM+NVM} memory model, especially when compared to NVM-only architectures. 
Eq. \ref{eq:energy-sram+NVM} is therefore enhanced by considering the leakage power of low-power SRAMs of the appropriate size, as reported in Table~\ref{tab:mem-energy-values}. 
The leakage power of STT and RRAM is considered to be zero, which is obviously not the case for real designs.
Table \ref{tab:bench-energy-values} reports for each benchmark the absolute dynamic energy of memory accesses for Freezer with both RRAM and STT as NVMs, equivalent to the Freezer blue bar in Figures \ref{fig:rram} and \ref{fig:stt}, respectively.
The table also reports an estimation of the leakage energy due to the main SRAM memory obtained considering a $20MHz$ clock, and the total memory size of the benchmark.
Table \ref{tab:bench-energy-values} shows that the leakage energy represents around half of the dynamic energy of memory accesses when using Freezer.
Even accounting for the leakage of SRAM, the approaches based on SRAM+NVMs are still better than running an NVM-only system.
Compared to full-memory backup which would consume roughly the same leakage energy, Freezer still benefits from the backup size reduction.

Moreover, even accounting for the leakage of the NVM memories would not change the outcome of the analysis.
In fact, when considering NVMs of the same size running for similar periods of time, the leakage due to the NVMs would be roughly the same for both SRAM+NVM and NVM-only architectures.
Furthermore, an SRAM+NMV system would even be able to activate the NVM only during the backup and restore phases, reducing even more the impact of NVM leakage.  
In both cases, the SRAM+NVM architecture would still show an advantage.
%

\begin{table}
    \centering
    \caption{{Backup energy using Freezer, leakage and memory size for different benchmarks, energy in $[\mu J]$, memory size in words of 32 bits.}}
    \begin{tabular}{lrrrr}
    \hline
    Trace &  mem\_size &  E\_freezer &  E\_freezer &  E\_leakage \\
    ~  & [32-bit word] &        RRAM &         STT &        SRAM \\
    \hline
      sss &      8192  &        11.0 &        11.0 &      6.1    \\
      ses &     16384  &         3.3 &         3.2 &      1.9    \\
    mm16f &      2048  &         2.5 &         2.4 &      2.0    \\
      fft &      2048  &         3.4 &         3.4 &      3.7    \\
    cjpeg &      8192  &         3.9 &         3.6 &      1.3    \\
       sl &      8192  &         3.6 &         3.7 &      1.9    \\
    mm16i &      1024  &       0.051 &       0.045 &    0.028    \\
     dijk &     16384  &        51.0 &        50.0 &     27.0    \\
    mm32i &      4096  &        0.58 &        0.56 &     0.41    \\
    \hline
    \end{tabular}
    \label{tab:bench-energy-values}
\end{table}

\subsection{Energy and Area Overhead Considerations}
\label{sec:area}

In this section, we provide insights about the overhead in energy due to our backup controller.
The use of the Freezer hardware backup strategy in an energy harvesting platform will introduce a small overhead at run-time, but will also decrease the energy required for the backup and restore operations. 
We can account for the overhead and the reduction in the backup size by modifying Eq. \ref{eq:energy} which becomes
\begin{equation}
    E_{c} = E_{s} N'_{s} + E_{r} N_{r} + (P_{on}+P_{ovh})\times t_{on} + P_{off} t_{off},
\end{equation}
where $N'_s$ is the reduced backup size and $P_{ovh}$ represents the overhead introduced at run-time.
The energy required for moving the data ($E_{s}$ for save and $E_{r}$ for restore) is heavily dependent on the memory technology.
However, software-based approaches introduces additional overhead.
In our case, as a backup operation may require hundreds or even thousands of transfers, we can approximate the energy required for saving one word as
\begin{equation}
E_{s} = E_{sram/r} + E_{nvm/w}
\label{eq:es}
\end{equation}
where $E_{sram/r}$ is the energy for reading a from the SRAM and $E_{nvm/w}$ the energy required for a write in the NVM.

The power overhead introduced by our strategy can be estimated as  $P_{ovh} = \alpha \times P_{active} + P_{leak}$, where $P_{leak}$ is the leakage power, which will be mostly determined by the \textit{to\_backup} memory, and $P_{active}$ the active power.
$P_{leak}$ and $P_{active}$ were provided in Section \ref{sec:areapowefreezer}.
$P_{active}$ will be consumed whenever the processor performs a store operation and $\alpha = N_{store}/N_{prog}$ is the fraction of clock cycles spent performing store operations w.r.t. the execution of the program in the whole interval.

This overhead can be compared with the advantage gained in terms of save and restore energy.
If we compare against a system that saves everything but does not introduce any overhead, we can estimate the maximum active time $t_{on}$ after which the power consumed by the controller during active time becomes greater than the energy reduction obtained at backup time. 
$t_{on}$ is constrained by the following inequation:
\begin{equation}
    t_{on} \le \frac{\delta E_{s} N_{tot}}{P_{ovh}}
    \label{eq:ton}
\end{equation}
where $N_{tot}$ is the number of words to be backed-up without Freezer (full memory), $E_{s}$ the energy required to back-up one word ($E_{sram/r} + E_{nvm/w}$), and $\delta E_{s} N_{tot}$ the energy saved during the backup operation.
With Freezer, considering $\delta = 87.7\%$, $E_{sram/r} = 0.45 pJ/bit$, $E_{nvm/w} = 100\times E_{sram}$, we obtained for the two extreme configurations depending on the considered benchmark:

$t_{on}~<~16.42s$ and $P_{ovh}= 1.18 \mu W$ for \textit{susan\_smooth}, and

$t_{on} < 2.4 s$ and a similar $P_{ovh}$ for the \textit{FFT} benchmark.\\
Both these $t_{on}$ values allow for the programs to be executed completely and are well above the typical active time of intermittently-powered systems.
Moreover, Eq. \ref{eq:ton} is obtained by comparing our solution to a system that introduces no overhead at run-time and no overhead during the backup process, which would not be the case in real systems. \\

To give an idea of how Freezer would fit in a low-end IoT node, we can compare it with a ultra-low-power, size-optimised SoC, implemented with the same 28nm FDSOI technology node such as the one presented in \cite{bol_a_40_to_80mhz_2019}.
In terms of area the SoC is 0.7$mm^2$, while its power consumption is 3$\mu$W/MHz giving at 48MHz a power consumption of 144$\mu$W.
From these numbers we can see that Freezer, even with our non-optimised implementation, would lead to a small overhead.
In particular, assuming blocks of 8 words, the area overhead of $2,748\mu m^2$ represents $\approx 0.4\%$.
The power overhead during active time, considering the $\alpha$ of the FFT benchmark, could be as low as 0.82\%.


\section{Discussion About The Approach}\label{sec:discussion}
Several studies have approached the problem of computing under intermittent power supply, providing a wide variety of different solutions. 
While software-based approaches try to solve the problem at the application level, hardware-based solutions try to provide platforms that implement the non-volatility in a way that is transparent to the programmer.
The majority of the hardware solutions usually rely heavily on the underlying memory technology to accomplish the state retention.
Even in \cite{hager_a_scan-chain:2017}, where no NVM is used, their technique relies on an ultra low-power retention SRAM.

Our approach moves away from this type of scheme and tries to solve the problem from a different standpoint, by providing hardware acceleration for the backup and restore procedures, and by exploiting run-time information to optimize the backup sequence.
Moreover, this approach is agnostic with respect to the NVM technology, and opens a series of possibilities.
Technologies such as hybrid nvSRAM, as the one used in \cite{liu_4.7_2016}, with circuit-level configurable memory, parallel block-wise backup and adaptive restore, may be exploited and enhanced by Freezer, thus achieving a faster and more energy efficient backup sequence thanks to the backup size reduction.

Furthermore, our approach could be extended to implement a programmable backup hardware accelerator, or to implement a dedicated ISA extension.
This would provide programs with some levels of control on the save and restore procedures and allow for the hardware to exploit some of the information available to the program.
As an example, a program may signal that a certain buffer or memory region is no longer used, allowing the controller to exclude it from the backup process.
This would also make possible to integrate static analysis techniques such as the one presented in \cite{zhao_software_2015} and \cite{zhao_stack-size_2017} on top of Freezer.

\section{Conclusion}\label{sec:conclusion}
Applications that run under ambient harvested energy suffer from frequent and unpredictable power losses.
To guarantee progress of computation in this circumstances, these applications have to rely on some mechanisms to retain their state.
In this paper, we propose Freezer, a backup and restore controller that is able to reduce the backup size by monitoring the memory accesses, and that provides hardware acceleration for the backup and restore procedures.
The controller only requires a small memory to keep track of the store operations.
Moreover, it can be implemented with plain CMOS technology and does not rely on complex and expensive hybrid non-volatile memory elements.
Furthermore, Freezer is a drop-in component that can be integrated in existing SoCs without requiring modifications to the internal architecture of the processor.
Our proposed solution achieve a 87.7\% average reduction in backup size on a set of benchmarks, and a two orders of magnitude reduction in the backup time when compared with software based state-of-the-art approaches.

The code and traces used in this paper are available for reproducibility at \url{https://gitlab.inria.fr/dpala/freezer-resources}.

\section*{Acknowledgment}
This work was supported by Inria Project Lab “Zero-Power Systems” (ZEP).
The authors would like to thank the anonymous reviewers for their comments and feedback.



\bibliographystyle{IEEEtran}
\bibliography{bibliography.bib}


\end{document}